\documentclass[useAMS,usenatbib]{mn2e}

\usepackage{graphicx}
\graphicspath{{./figures/}}

 \usepackage[T1]{fontenc} 
 \usepackage{aecompl}
\usepackage{aas_macros}

\title[Star formation properties of sub-mJy radio sources]{Star formation properties of sub-mJy radio sources}

\author[Bonzini et al.]{M. Bonzini\thanks{E-mail:marghe.bonzini@gmail.com}$^1$ 
V. Mainieri$^1$, P. Padovani$^1$,
P. Andreani$^1$, S. Berta$^2$, M. Bethermin$^1$,
\newauthor % S. Vattakunnel$^4$,
  D. Lutz$^2$, G. Rodighiero$^3$, D. Rosario$^2$, 
 P. Tozzi$^4$, S. Vattakunnel$^5$
\\
$^1$ European Southern Observatory, Karl-Schwarzschild-Strasse 2, D--85748 Garching, Germany\\
$^2$ Max-Planck-Institut f\"ur extraterrestrische Physik, Giessenbachstra\ss{}e, D-85748 Garching, Germany\\
$^3$ Dipartimento di Fisica e Astronomia, Universit\`a di Padova, vicolo dell'Osservatorio 3, I--35122 Padova, Italy\\
$^4$ INAF - Osservatorio Astrofisico di Arcetri, Largo E. Fermi, I-50125, Firenze, Italy\\
$^5$ INAF - Osservatorio Astronomico di Trieste, via G.B. Tiepolo 11, I-34131, Trieste, Italy
}

\begin{document}

\date{July 2015}

\pagerange{\pageref{firstpage}--\pageref{lastpage}}  \pubyear{2015}

\maketitle

\label{firstpage}
\begin{abstract}
We investigate the star formation properties of $\sim$ 800 sources detected in one of the deepest radio surveys 
at 1.4 GHz. 
Our sample spans a wide redshift range ($\sim 0.1 - 4$) and about four orders of magnitude in star formation rate (SFR). It includes both star forming galaxies (SFGs) and active galactic nuclei (AGNs), further divided into radio-quiet and radio-loud objects.
We compare the SFR derived from the far infrared luminosity, as traced by \textit{Herschel}, with the SFR computed from their radio emission. 
We find that the radio power is a good SFR tracer not only for pure SFGs but also in the host galaxies of RQ AGNs, with no significant deviation with redshift or specific SFR.  
Moreover, we quantify the contribution of the starburst activity in the SFGs population and the occurrence of AGNs in sources with different level of star formation.
Finally we discuss the possibility of using deep radio survey as a tool to study the cosmic star formation history. 
\end{abstract}

\begin{keywords}
galaxies: active --- galaxies: starburst --- galaxies: star formation --- radio continuum: galaxies --- surveys
\end{keywords}

\section{Introduction}
\label{intro}
The faint radio sky is a complex mixture of star forming galaxies (SFG) and active galactic nuclei (AGN) \citep[e.g.,][]{mauch07,smolcic08,padovani09,padovani11b}. 
Indeed, radio emission can either be due to the relativistic jets powered by the AGN or to synchrotron emission from electrons accelerated by supernova explosions.
In the latter case, radio emission can then be used as a star formation rate (SFR) tracer \citep[e.g.,][]{yun01,pannella09} which opens the possibility of using radio survey to study the cosmic star formation history (CSFH). 
Moreover, the radio band has the advantage, compared to UV or optical frequencies, to offer a dust unbiased estimate of the SFR being almost unaffected by dust extinction.

However, the majority of radio surveys available up to now were only sensitive to the most powerful radio loud (RL) AGNs, whose radio emission is mainly due to the relativistic jets rather than to star formation. In the few cases in which the sensitivity was high enough to detect SFGs, only the most extreme starburst galaxies, with SFRs $\ga 1,000~M_{\odot}$ yr$^{-1}$, could be detected above redshift about 1, at the mJy level. It is only going below 0.1 mJy, that we become sensitive, over a wider redshift range, to the bulk of the SFG population, which moreover represents the dominant contribution to the overall radio population at these radio flux densities \citep[e.g.,][]{padovani11b,bonzini13,padovani15}.
Such a sensitivity has been reached so far
only on small patches of the sky as with the VLA survey of the Extended Chandra Deep Field South (E-CDFS) considered in this work. 

A further problem that has limited so far the use of radio surveys for studying the CSFH is the challenge of separating the two radio emission mechanisms, jets and star formation, in faint radio samples.
%in particular from radio-quiet (RQ) AGNs. Indeed, while RL AGNs can be relatively easily distinguished from SFGs \citep[see][for details]{bonzini13}, RQ AGNs can have host galaxy properties very similar to SFGs, especially in type II objects where the AGN emission is obscured by dust \citep{bonzini13}. 
In particular, the origin of radio emission in the so-called radio-quiet (RQ) AGNs has been a matter of debate 
for quite some time. It has been proposed that these sources are a scaled down version
of RL objects \citep[e.g.,][]{miller03,giroletti09} or that the radio emission is mostly due to
the star formation in the host galaxy\citep[e.g.,][]{sopp91}. 
In the former case, the SFR derived from the radio power would be overestimated compared to the one obtained by other SFR tracers like, for example, the far-infrared (FIR) luminosity.

If instead radio emission in RQ AGNs is due to SF, the radio power could be used as an SFR tracer even when the AGN emission strongly contaminates the optical-to-MIR host galaxy light. 
This would allow us to include powerful AGN hosts in SFR studies therefore superseding the limitation of many current works that are restricted to pure star forming systems. However, it is first necessary to find an effective method to separate RQ and RL AGNs, as in the latter sources the radio luminosity primarily traces the jets emission. In \citet{bonzini13} we have proposed a method to separate these two AGN populations. We note that the RL AGNs contribution to the cosmic star formation history is expected to be low for two main reasons; firstly, RL AGNs represent only a small fraction (10\%) of the radio sources at flux densities of $\sim30\ \mu$Jy and, secondly, their host galaxies have usually low star formation rates as they are mainly passive systems.
%These bright AGNs are usually excluded a priori from SFR studies leading to a biased view of the actively star-forming population.

Moreover, estimating the SF activity in AGNs host galaxies is particularly important to investigate the possible impact of the AGN in shaping its host galaxy properties.
Indeed, both from the theoretical and observational side, there has been a lot of debate on the role the AGN can have in regulating the SF activity, either triggering or suppressing it \citep[e.g., ][]{mullaney11,rosario12,page12,harrison12,zubovas13}.

In this work we exploit the star-formation properties of a large sample of radio sources selected in a deep VLA survey of the E-CDFS down to a 5$\sigma$ flux density limit of about 32 $\mu$Jy. 
The paper is organized as follows. In section \ref{sec_sample} we described our radio sample, how it splits in the different source populations, and the ancillary data used for the analysis. 
After describing the method adopted to estimate the FIR luminosity of the radio sources (Sec. \ref{sec_Lfir}), we present the radio-FIR correlation for radio selected SFGs (Sec. \ref{sec_RFC}). 
In Section \ref{sec_SFR} we compute the SFR from the radio and FIR luminosity and we compare them in Sec. \ref{sec_radio_in_RQ}. The position of our radio selected sources in the SFR-stellar mass plane is described in Sec. \ref{sec_SFR_vs_Mstar} and in Sec. \ref{sec_sSFR} their SF activity is investigated. 
In Sec. \ref{sec_discussion} and \ref{sec_summary}, we discuss and summarize our results, respectively. 
Finally, Appendix \ref{sec_model_param} is dedicated to the description of the empirical model we compare with our observations and in Appendix \ref{sec_cat} we describe the catalogue with the physical properties of the radio sources used in this work that we make publicly available.  

In this paper we assume a cosmology with $H_{0}=70$ km s$^{-1}$
Mpc$^{-1}$, $\Omega_{M}=0.27$, and $\Omega_{\Lambda}=0.73$ and a Chabrier initial mass function \citep[IMF;][]{chabrier03}.

%____________________________________________________________
\section{Sample description}
\label{sec_sample}
\subsection{Radio data}
The Extended Chandra Deep Field South (E-CDFS) has been observed with
the Very Large Array (VLA) at 1.4 GHz between June and September 2007
\citep{miller08}. The survey reaches a best rms sensitivity of
6 $\mu$Jy and the average 5$\sigma$ flux density limit is 37 $\mu$Jy
with near-uniform coverage.
%the spatial resolution is $2.8\arcsec \times 1.6\arcsec$.
A catalogue including sources down to peak flux density of five times
the local rms noise has been extracted.  A description of the survey
strategy and the data reduction details are given in \citet{miller13}.
The radio catalogue includes 883 sources. Using a likelihood ratio
technique \citep{ciliegi03}, the optical/infrared counterparts of the radio sources have
been identified \citep{bonzini12}. Excluding the outermost region as
defined in \citet{bonzini12}, the wealth of multi-wavelength data
available for the E-CDFS allows robust estimates of photometric
redshifts \citep{santini09,taylor09,cardamone10,rafferty11}. The area
with photometric redshifts coverage includes 779 radio sources that
will constitute our main sample for the analysis presented in this
paper. Combining the photometric redshifts with the output of several
spectroscopic campaign in the E-CDFS, we were able to assign a
redshift to a total of 675 radio sources, 37\% of which are
spectroscopic \citep[see][for details]{bonzini12}. The average redshift is $\langle z \rangle \sim 1.1$.

\subsection{Radio source populations}
\label{sec_jets_or_SF}
As mentioned in Section \ref{intro}, the sub-mJy radio population is a
mixture of SFG and AGN. The latter further divides into RQ and RL AGN.
To separate these different populations we use a multi-wavelength
approach combining radio, mid-infrared, and X-ray data. A detailed
description of our classification scheme is given in
\citet{bonzini13}; here we briefly summarize its main characteristics.
We select RL AGN using the $q_{24obs}$ parameter, which
is the logarithm of the ratio between the observed 24$~\mu$m flux
density and the observed 1.4 GHz flux density \citep[e.g.,][and references therein]{sargent10}. In the $q_{24obs}$-redshift plane we have defined an
``SFG locus'' below which we find radio sources with a radio excess compared to the typical $q_{24obs}$ ratio of SF systems. This excess is the signature of an
AGN contribution to the radio luminosity; these sources have been referred to as radio-dominant AGNs \citep{appleton04}, radio-excess AGNs \citep[e.g.][]{drake03,delmoro13} or simply radio-loud AGNs \citep[e.g.][]{donley05,padovani09} in the literature. Here, we adopt the nomenclature RL AGNs.
The SFG locus has been obtained computing the $q_{24obs}$ as a function of redshift for the star forming galaxy M82 \citep{polletta07} and assuming a 0.7 dex dispersion. In \citet{bonzini13} we have discussed the implications  of choosing the M82 template for source classification (see also \ref{sec_radio_in_RQ}). 
Within and above this locus, both SFGs and RQ AGNs can be found \citep[e.g.][]{donley05, bonzini13}. A source is classified as RQ AGN if there is any evidence of AGN activity in the other bands considered: we classify a radio source as RQ AGN if a) it has a hard band [2-10 keV] X-ray luminosity greater than $10^{42}$ erg s$^{-1}$; b) it lies in the ``AGN wedge'' of the IRAC color-color diagram, as defined by \citet{Donley12}. Otherwise, the object is classified as a SFG. 
According to this scheme, our sample of 779 radio sources
includes 167 RL AGNs, 188 RQ AGNs, and 424 SFGs.

\subsection{\textit{Herschel} data} 
\label{sec_herschel_data}
The E-CDFS has been observed by \textit{Herschel} as part of the PEP
(PACS Evolutionary Probe) programme \citep{lutz11}. The
\textit{Herschel} observations are deeper in the central part of the
field, the GOODS field, and shallower in the outskirts.
The whole
field has been observed at 100 and 160 $\mu$m and for the central part
70$\mu$m data are also available. For GOODS proper we use the combined reduction of PEP and GOODS-\textit{Herschel} \citep{elbaz11} PACS data as described in \citet{magnelli13}. The 5$\sigma$ flux level of the 100
(160) $\mu$m maps are 0.85 (2.1) mJy and 6.25 (13.05) mJy in the GOODS
and E-CDFS, respectively. The 70 $\mu$m 5$\sigma$ flux density limit
is 1.35 mJy. \textit{Herschel} photometry was performed through point
spread function (PSF) fitting, adopting Spitzer MIPS 24 $\mu$m
detected sources as positional priors \citep{berta13}.  Blind catalogues
are also extracted by means of PSF-fitting using the StarFinder IDL
code \citep{diolaiti00a, diolaiti00b} and include all sources with
$S/N>3\sigma$.  We cross-correlate this \textit{Herschel}/PACS
catalogues with our radio sample using a 1.5$\arcsec$ searching radius.
We find a total of 490 matches, 33\% in the GOODS field and 67\% in
the outer region of the E-CDFS.

In particular, in the PACS-radio sub-sample we have 42 RL AGN, 130 RQ
AGN, and 311 SFG. We note that in the PACS sample we recover 69\% of
the RQ AGN, 78\% of the SFG but only 25\% of the RL AGN from the
original radio selected sample considered in this paper. The small
fraction of RL AGN is not surprising since they are preferentially
hosted in passive, dust poor, galaxies. We therefore expect only a small fraction of them to be detected in the FIR.

%Upper limits
For the radio sources without a PACS counterpart we compute 5 $\sigma$
upper limits from the local rms noise, therefore taking into account
the non uniform coverage of the \textit{Herschel} observations in this
field.

\section{FIR luminosity of radio sources}
\label{sec_Lfir}
The \textit{Herschel} photometry is crucial to have a good estimate of
the FIR emission since it allows to trace the cold dust
emission. At a typical redshift of $\sim
1$, the PACS measurements probe the rising part and/or the peak of the rest-frame FIR dust
emission bump. We apply a fitting technique to the full
UV-to-FIR spectral energy distribution (SED) to better constrain the
FIR luminosity of our radio sample, taking advantage of the
exquisite multi-wavelength coverage available in the E-CDFS.

The UV-NIR photometry is obtained combining the BVR selected
\citet{cardamone10} catalogue, the K selected \citet{taylor09}
catalogue, and the \citet{damen11} IRAC selected SIMPLE catalogue. In the MIR, we used the MIPS 24
and 70$\mu$m observations from the Far-Infrared Deep Extragalactic
Legacy Survey (FIDEL) \citep{dickinson07}. A likelihood ratio technique \citep{ciliegi03} followed by a careful visual inspection has been used to reconstruct the optical-to-MIR photometry of the radio sources \citep[see][for details]{bonzini13}. At longer wavelengths, given the lower surface density of sources, we simply cross-correlate the \textit{Herschel} and radio counterparts catalogues, as described in the previous
section. In the GOODS field, the PACS 70$\mu$m detection rather than
the MIPS one is adopted when available. In summary, for each radio
source we have on average 16 photometric points.

%%Library
To fit the UV-to-FIR photometry we use the \citet{berta13} template
library that is based on the observed SED of PACS detected sources. It
is a collection of 32 templates, mostly SFGs, of which twelve also
include an AGN component.
%In 2/3 of them the contribution of the AGN to the FIR luminosity less than 10\%

We chose this library mostly for its wide range of SED shapes. This
is important if we want to fit the photometry of our radio
selected sample since, as already mentioned, it is highly
non-homogeneous: it includes both normal star forming and starburst
galaxies, objects with the clear presence of a powerful AGN and
sources whose SED is dominated at all wavelengths by the emission from
stars. However, as the \citet{berta13} library is based on PACS
detected sources, it lacks galaxies with very old stellar
population and consequently very low emission in the FIR. Therefore, we decided to
add three templates of elliptical galaxies from the SWIRE template
library \citep{polletta07}.

\begin{figure*}
\begin{tabular}{c c c}
	\includegraphics[trim=0.5cm 0.3cm 0.8cm 0.4cm, clip=true,width=0.65\columnwidth]{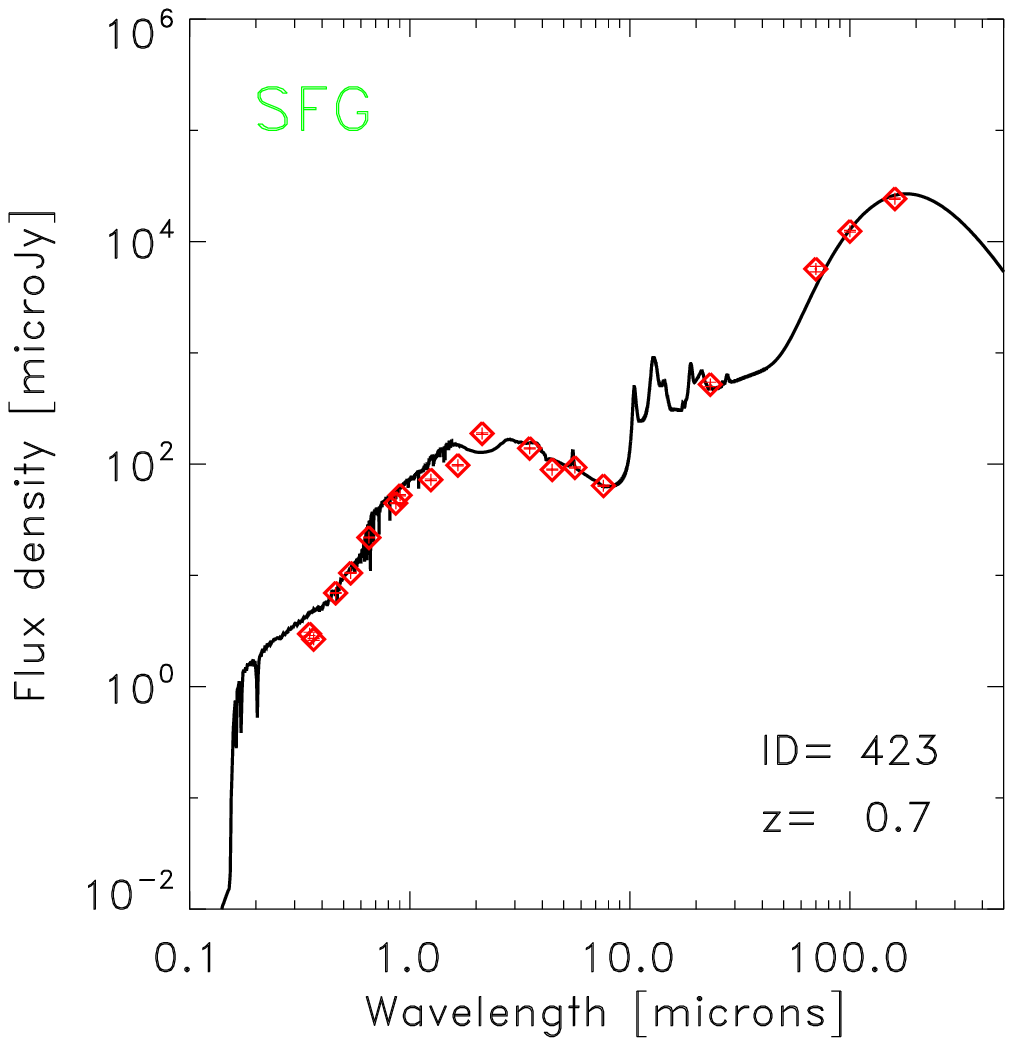} &
	\includegraphics[trim=0.5cm 0.3cm 0.8cm 0.4cm, clip=true,width=0.65\columnwidth]{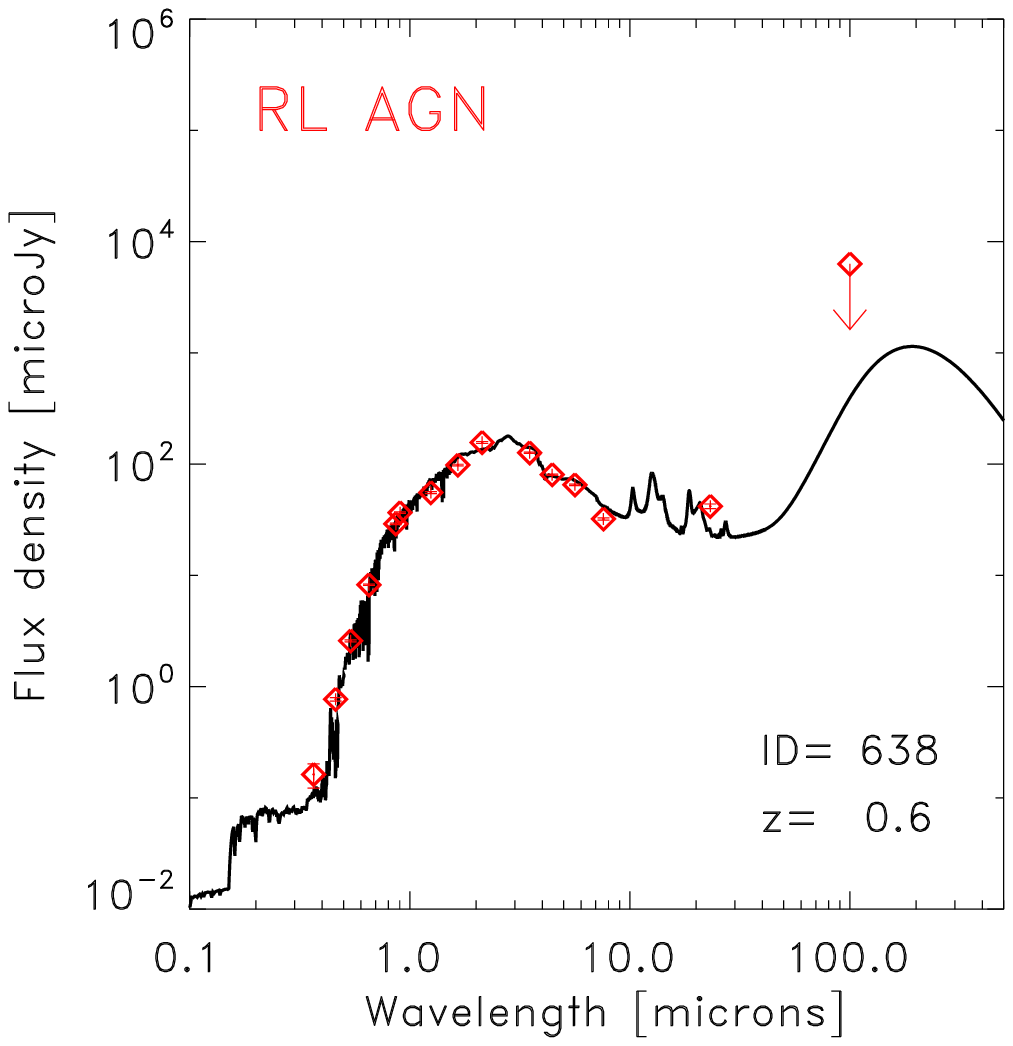} &
	\includegraphics[trim=0.5cm 0.3cm 0.8cm 0.4cm, clip=true,width=0.65\columnwidth]{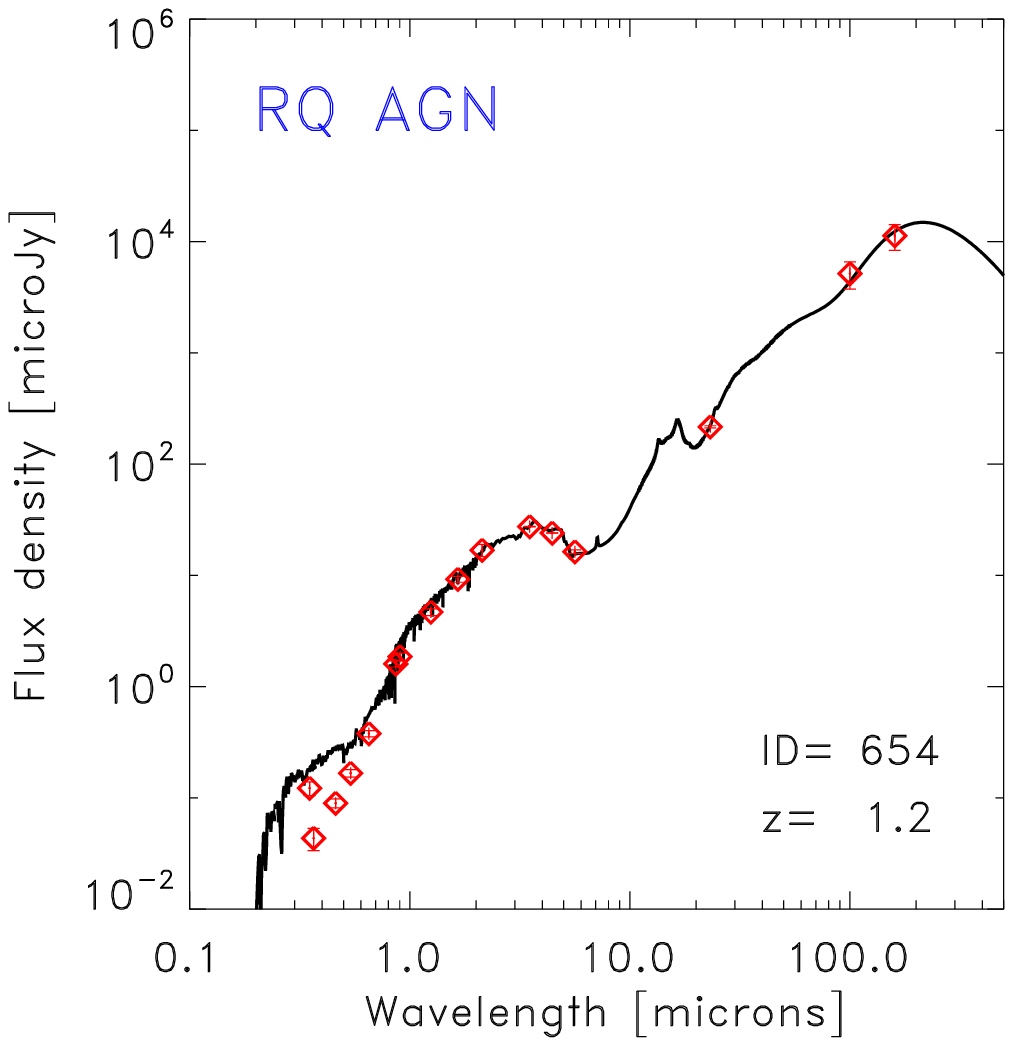} \\

\end{tabular}	
\centering

 \caption[Examples of the optical-to-FIR SED fitting.]{\small{Examples of the optical-to-FIR SED fitting for the three class of sources. The combined photometry is plotted as red symbols as a function of wavelength (observed frame) and the solid line show the best fit template. Source ID and redshift are given in the bottom right corner.}}
 \label{fig_SEDs}
\end{figure*}

%%Method
We use a standard $\chi^2$ minimization technique to find the best fit template. Examples of the optical-to-FIR SED fitting for the three class of sources are shown in Fig. \ref{fig_SEDs}. For sources without \textit{Spitzer} and/or \textit{Herschel} detections we imposed the best fit template to not exceed the 5 $\sigma$ upper limits in these bands. We finally compute the total FIR luminosity ($L_{\rm FIR}$) integrating the best
fit template between 8 and 1000$\mu$m.
Errors are computed repeating the fitting procedure 100 times for each source,
randomly modifying the photometry within the uncertainties and then
taking the standard deviation of the $L_{\rm FIR}$ distribution
obtained.

We note that for sources without PACS detection the FIR luminosity is
derived from the best fit based only on the optical-MIR photometry,
taking into account the upper limits in the FIR. These luminosity
estimates are therefore less robust than the one constrained by the
PACS measurements. 

\section{Radio-FIR correlation}
\label{sec_RFC}
The radio and FIR luminosity in SFGs follow a tight empirical relation, the so-called "radio--FIR
correlation" (RFC) \citep[e.g.,][]{dickey84,dejong85,bell03}. 

The physical reason for its tightness and linearity over several decades in luminosity has been investigated by several theoretical models \citep[e.g.,][]{lacki10a,lacki10b,schleicher13}, but remains still unclear. These works also predict that the RFC should break at high $z$ where inverse Compton losses start to dominate over the non-thermal synchrotron emission.

From the observational point of view, there are some hints of possible deviations at high z or in extreme objects like sub-millimetre galaxies, nevertheless these observations are still consistent with no significant evolution \citep[e.g.,][and reference therein]{ivison10a,moric10,sargent10,magnelli12a,lutz14}.

Deep radio surveys would be needed to investigate the behaviour of the RFC at high redshift for normal SFGs. Indeed, even to detect star forming galaxies with
SFR of hundreds of solar masses per year at z$>$2, a $\mu$Jy radio
sensitivity is required.  Such a sensitivity has been reached so far
only for small patches of the sky like in the VLA survey of the E-CDFS
considered in this work. 

\begin{figure}
	\centering
	\includegraphics[trim=0.3cm 0.6cm 0.1cm 0.1cm, clip=true,width=\columnwidth]{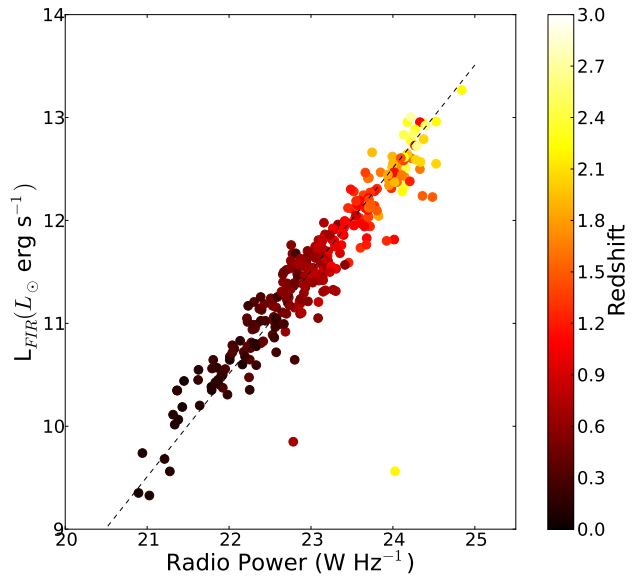}
 \caption[Radio power versus FIR luminosity for SFGs detected by \textit{Herschel}.]{\small{Radio power  at 1.4 GHz versus FIR luminosity for SFGs detected by \textit{Herschel}. The color scale represent the redshift of the sources and the dashed line shows the \citet{kennicutt98a} relation given in Eq. \ref{eq_RFC}}.}
 \label{fig_Pr_vs_Lfir_SFG_z}
\end{figure}

In this section we therefore investigate the
radio-FIR correlation for our radio selected SFGs up to z$\sim$ 3.  In
Figure \ref{fig_Pr_vs_Lfir_SFG_z} we show the radio power versus the
FIR luminosity for SFGs with a clear detection in at least one PACS
filter since their FIR luminosity estimate are more robust.
With the exception of two clear outliers\footnote{RID 521 (dark point) is a resolved edge-on spiral whose 24$\mu$m emission has been probably underestimated due to aperture photometry. As a consequence the fit is poor and the FIR luminosity is underestimated. RID 577 (light point): there is a small offset between the radio and optical-MIR emission hence it is possible that the radio emission is associated with a background object.}, 
the two luminosities are tightly correlated and lie along the empirical RFC \citep[e.g.,][]{kennicutt98a} plotted as dot-dashed line and parametrized by:
\begin{equation}
\label{eq_RFC}
\log(P_{1.4GHz})=\log(L_{\rm FIR}) + 11.47 
\end{equation}
%or 11.47
where $P_{1.4GHz}$ is the radio power in W Hz$^{-1}$ at 1.4 GHz and $L_{\rm FIR}$ is the
FIR luminosity in unit of solar luminosity expressed in erg s$^{-1}$.
The average dispersion is 0.2 dex. We note that such a dispersion is smaller compared to the dispersion in $q_{24obs}$ values for our SFGs that is 0.33 dex. This implies that the correlation between the total FIR luminosity and the radio power is stronger that the one obtained using only MIR data.
The color scale in
Fig. \ref{fig_Pr_vs_Lfir_SFG_z} represents the redshift of the
sources. Thanks to the $\mu$Jy sensitivity of the VLA observation, we
can detect SFGs in a wide redshift range, from z$\sim$0.1 to z$\sim$ 3
and we find that the RFC holds over the whole redshift range with
almost constant dispersion.
Moreover, our data probe four orders of magnitudes in luminosity and 
therefore include both normal SFG as well as the most active systems with SFR
of a thousand of solar masses per year. The SF in these latter systems
is thought to be triggered by episodic violent events such as major
mergers rather than by secular processes as in normal SFGs
\citep[e.g.,][]{daddi10b}. Our study suggests that the correlation
between radio and FIR emission is the same in both kinds of systems, at least down to the luminosities probed by our observations. 
However, we still cannot exclude a breakdown at lower star formation surface densities as predicted by e.g. \citet{schleicher13} for which even deeper radio data would be needed.  

This results justify the use of both the FIR and radio luminosity as
independent but equivalent SFR tracers for our radio selected SFGs.
It will be a useful element in our investigation on the origin of the
radio emission in RQ AGN (see Sec. \ref{sec_radio_in_RQ}).

\section{Star formation rate estimate}
\label{sec_SFR}
The FIR luminosity is a tracer of recent SF activity as the UV
radiation from young stars is absorbed by the dust and re-emitted at
longer wavelengths, in the FIR.  The $L_{\rm FIR}$ provides a better
estimate of the SFR compared to the one derived from the UV or line
emission since these measurements are highly affected by dust
extinction. Moreover, computing the SFR from the FIR is particularly
useful for bright AGN hosts. Indeed, the AGN emission can overshine the
emission of the whole galaxy at optical wavelength, while it
represents a tiny fraction ($<$10\%) of the total luminosity in the
FIR in most of the cases \citep[e.g.,][]{hatziminaoglou10,rosario12,berta13}. %[see references in Rosario+12, Sec3.1].
We adopt the \citet{kennicutt98a} relation to compute the SFR from the FIR luminosity:
\begin{equation}
SFR_{\rm FIR}[M_{\odot}\ yr^{-1}]= 4.5 \times 10^{-44} L_{\rm FIR} [erg  s^{-1}].
\end{equation}
As this equation assumes a Salpeter IMF, we rescaled it for a Chabrier IMF multiplying by a factor of 0.6. 
 
%SFR from the radio
Also the radio continuum emission can be used as a tracer of recent star formation in SFGs since it is
nearly all due to synchrotron emission from relativistic electrons
associated to SN remnants \citep{condon92}. The empirical conversion between the radio power at 1.4
GHz and the SFR of the galaxy according to
\citet{yun01} is:
%pannella11.
\begin{equation}
\label{eq_sfr_radio}
SFR_{\rm r}[M_{\sun}\ yr^{-1}]=(5.9 \pm 1.8) \times 10^{-22} P_{1.4GHz} [W Hz^{-1}].
\end{equation}
We multiply the derived SFR by a factor of 0.6 to convert from a Salpeter to
a Chabrier IMF.

\section{Radio emission in RQ AGN} 
\label{sec_radio_in_RQ} 
\begin{figure}
	\centering
	\includegraphics[width=\columnwidth]{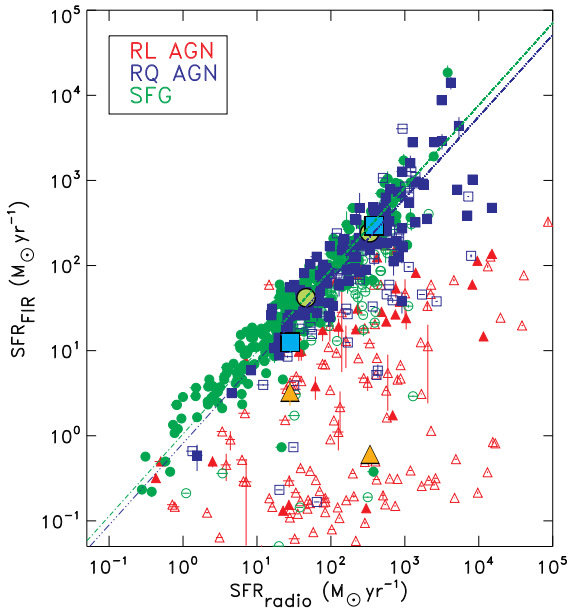}
 \caption[SFR derived from the FIR luminosity versus the SFR
     from the radio luminosity.]{\small{SFR derived from the FIR luminosity versus the SFR
     from the radio power at 1.4 GHz. SFGs are plotted as green circles, RQ
     AGNs as blue squares and RL AGNs as red triangles. Full symbols
     represent sources detected in at least one PACS filter, while
     sources shown as empty symbols are \textit{Herschel}
     non-detections. Large symbols with lighter colors are the results of the stacking
     analysis (see Section \ref{sec_stacking} for details).}}
 \label{fig_sfr_comparison}
\end{figure}

In Figure \ref{fig_sfr_comparison} we compare the SFR computed from
the FIR luminosity with the SFR derived from the radio luminosity for
all the 675 radio sources with redshift. Different colors represent
different classes of objects: green for SFGs, blue for RQ AGNs, and red for RL AGNs. Sources detected in at least one
\textit{Herschel} band are plotted with full symbols. We first
concentrate on these sources since their $SFR_{\rm FIR}$ is more
reliable.

As described in Section \ref{sec_RFC}, we find that the SFGs follow
the RFC over the whole range in luminosities accessible to our
survey. As a consequence, the two SFR estimates are in agreement over four
decades in SFR with a typical dispersion of 0.2 dex. More
interestingly, we find good agreement between the SFRs derived from the
two different tracers also for RQ AGN with only a slightly larger scatter of 0.23 dex.  The results of a linear
regression fit (least square bisector) are: % add ref
\begin{equation}
\log(SFR_{\rm FIR})=0.97\pm 0.02 \times \log(SFR_{\rm r}) + 0.02\pm 0.04 
\end{equation}
and
\begin{equation}
\log(SFR_{\rm FIR})=0.96\pm 0.05 \times \log(SFR_{\rm r}) - 0.09\pm 0.10 
\end{equation}
for SFGs and RQ AGNs with PACS detections, respectively. The two
relations are fully consistent within the uncertainties. This implies
that the radio power is a good tracer for the SFR not only for SFGs, but also for RQ
AGNs, with almost the same degree of accuracy for the two types of sources. We emphasize that this result is not a direct consequence of our selection method; indeed, even if both types of sources lie within the ``SFGs locus'' in the $q_{24obs}$-redshift plane they have different MIR characteristics. SFG, for example, have an average dispersion of $q_{24obs}$ values of only 0.33 dex while for RQ AGNs the dispersion is two times larger (0.65 dex). This is mainly due to the large AGN contribution at MIR wavelengths in many RQ AGNs. As a consequence, an estimate of the SFR from the observed 24$\mu$m emission would lead to a much more uncertain result for RQ AGNs as for SFGs. Instead, based on Fig. \ref{fig_sfr_comparison}, we can claim that the radio luminosity is as good tracer as the FIR of the star-formation in RQ AGN host galaxies.
The correlation shown in Fig. \ref{fig_sfr_comparison}, also strongly suggests that the main contribution to
the radio emission in RQ AGN is therefore due to the SF in the host
galaxy rather than black hole activity \citep[see also][]{padovani11b}. As mentioned above, that could not be claimed considering only MIR data as these wavelengths are not reliable for deriving the SFR in many powerful AGNs; it would have been impossible to resolve if the larger scatter in $q_{24obs}$ values for RQ AGNs compared to SFG was due to the presence of hot dust heated by the AGN or to the presence of small radio jets. Hence, it was crucial to include \textit{Herschel} data in our analysis that allowed us to obtain a reliable estimate of the SFR in our sources, assuming that the AGN contribution to the total FIR luminosity is in the large majority of the cases negligible \citep[][see also below]{hatziminaoglou10,rosario12,berta13}.

The behaviour of the RL AGNs further supports our claim that the origin of radio emission in the two types of AGN is different: they scatter
out from the one-to-one relation in Fig. \ref{fig_sfr_comparison} since
the SFR computed from the radio luminosity is overestimated due to the
jet contribution to the radio power.
Note that this is true not only for very powerful RL objects but also for low-power radio AGNs.
The comparison of the two SFR tracers allows us also to isolate sources that have been misclassified; indeed there are some RQ AGNs ($\sim 5\%$) with PACS detections that are below the relation, especially at high redshift. %(RID 216, 225, 508, 565, 639, 843) 
In \citet{bonzini13}, we estimated a contamination to the RQ AGNs population from RL AGNs with a strong contribution from the AGN to the 24 $\mu$m flux density that boosted their $q_{24obs}$ value into the SFGs locus of about 5\%, in agreement with our findings. 
%This effect is particularly strong at z$\sim$2.
Therefore, we believe that most of the outliers are indeed RL AGNs\footnote{We will keep the original classification in the rest of the paper. Nevertheless, we checked if our conclusions change reclassifying these sources, finding no significant differences.}.
These outliers are also the main responsible for the slightly larger scatter (0.23 dex) %computed as standard deviaton of the orthogonal distance from the best fit relation
of RQ AGNs as compared to the SFGs population (0.2 dex).

\subsection{AGN contribution in the FIR}
\label{sec_AGNinFIR}
As already mentioned in Sec. \ref{sec_Lfir}, twelve of the SED
templates in the library used to fit the photometry of our radio sample include
an AGN component. The AGN contribution to the FIR luminosity in these
templates has been estimated in \citet{berta13}. We have therefore
the opportunity to subtract the AGN contribution to the
$SFR_{\rm FIR}$ and analyse its impact on our conclusions. 
About 50\%
of the RQ AGNs and 35\% of the SFGs are best fitted by a template that
has an AGN component.
However, it is important to note that the fit solutions are highly degenerate; in particular, templates with small AGN contribution and pure SFG ones might provide equivalent good fits to the source photometry.
Hence, the results reported in the following are valid only in a statistical sense.
Nevertheless, the fact that only half of our RQ AGNs are best fitted by
an AGN template shows that a classification based only on the optical-to-FIR properties can miss a significant fraction of AGNs. For these latter sources, the AGN contribution to the optical-to-FIR SED is typically minor and, in most of the cases (80\%), they are identified as AGNs due to their high X-ray luminosity.  
On the other hand, the fact that we find some SFGs that are best
fitted by an AGN template does not mean that they are all real AGNs. As already highlighted, the SED fitting technique often yields degenerate solutions, especially in the case of discrete libraries. However, it is possible that some of these SFGs are indeed AGNs since our classification
scheme was aimed to select a clean sample of RQ AGNs and a
contamination from low-luminosity AGNs in SFG population is expected
\citep[see][for details]{bonzini13}.

For all the sources fitted with an AGN template we subtracted the AGN contribution to the FIR
luminosity and re-computed the corresponding SFR. The new best linear
regressions have a slope of 0.95 for both SFGs ($\pm 0.02$) and RQ
AGNs ($\pm 0.05$) and the offset from the 1:1 correlation are $0.04
\pm 0.05$ and $-0.10 \pm 0.11$, respectively. They are therefore
consistent with the previous estimates within the uncertainties. The
only difference is a slightly larger scatter around the best fit
relation of 0.21 dex and 0.25 dex for SFGs and RQ AGNs, respectively.

We conclude that the AGN contribution to the FIR emission as computed
from the best fit template has negligible impact on our results.
We will therefore consider the $SFR_{\rm FIR}$ non-corrected for AGN
contribution in the rest of this paper.

\subsection{Stacking of \textit{Herschel} undetected sources}
\label{sec_stacking}
\begin{figure}
	\centering
	\includegraphics[width=\columnwidth]{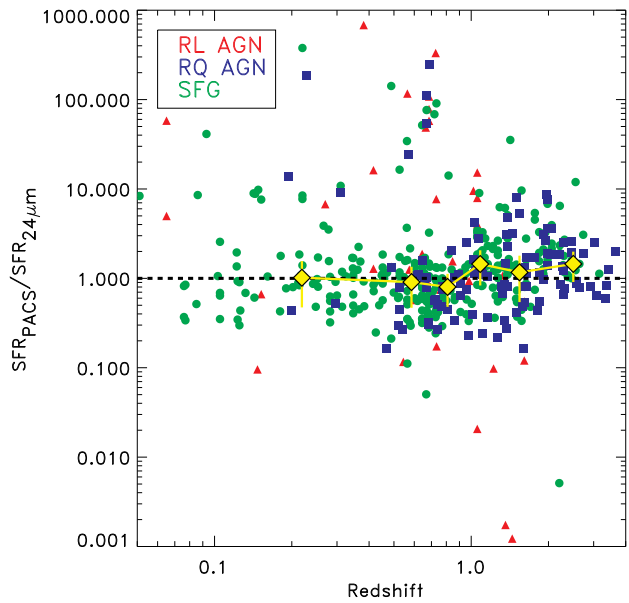}
 \caption[SFR from the full photometry and UV-to-MIR data ratio.]{\small{Ratio of the SFR computed from the fit to the whole
     photometry and the one obtained using only the UV-to-24 $\mu$m
     data as a function of redshift. Different colors correspond to
     different classes of sources as in
     Fig. \ref{fig_sfr_comparison}. Large yellow diamonds represent the median and the standard deviation in different redshift bins.}}
 \label{fig_SFRPACS-24_ratio_vs_z}
\end{figure}

As already discussed above, for sources without a PACS detection the
estimate of the FIR luminosity is less robust. Indeed, the best fit
template has been obtained considering the photometry only up to
24$\mu$m, with upper limits at longer wavelengths. To estimate how
this can impact the SFR estimate, we have re-computed, with the same
method, the SFR for sources \textit{with} PACS detection but
considering only their photometry up to 24 $\mu$m. In
Fig. \ref{fig_SFRPACS-24_ratio_vs_z} we plot the ratio of the two SFR
measurements, considering or not the PACS data, as a function of
redshift. Large yellow symbols represent the median and the standard deviation in different redshift bins.
We observe a large number of outliers ($\sim$25\%) and a slight tendency to underestimate the FIR emission especially above z$\sim$1 ($\sim0.16$ dex) when the \textit{Herschel} data are not included in the fitting procedure.

This test explains the larger scatter of the empty symbols in
Fig.\ref{fig_sfr_comparison} and justifies their exclusion from the
linear regression analysis.  However, it could be that the good
agreement we find between the two SFR tracers holds only for the
FIR-brightest objects while it breaks for the \textit{Herschel}
undetected sources.

Therefore, we performed a mean stacking analysis \citep{bethermin10} %cite also @@@ 
of the PACS-undetected sources to estimate their average $SFR_{\rm FIR}$.  We divide the undetected sources in classes and for each
class we split the sample into two bins of radio power ($21.5<P_{\rm r}<23.5$
and $23.5<P_{\rm r}<24.5$).  We then stacked both the 100 and 160$\mu$m
\textit{Herschel} maps at the position of the PACS-undetected radio
sources using the error maps \citep{lutz11} as weights. Flux densities are extracted
from the stacked images through PSF-fitting; aperture corrections and
correlated noise corrections are applied
\citep{berta13}. Uncertainties are estimated using a bootstrap
approach \citep{bethermin10} and corrected also for high-pass filtering
effects. Table \ref{tab_stacking} summarize the stacking analysis
results. We obtain $>2\sigma$ detections for SFG and RQ AGN in both
bands in each bin, with the only exception of the 100$\mu$m stacked
image for RQ AGN in the lower radio power bin. We note that this bin
has also the smallest number of sources (12) and therefore the S/N
ratio is lower. RL AGN are detected only at 100$\mu$m in the low
radio power bin while they are undetected ($<2\sigma$) elsewhere.
\begin{table*}
\small
\caption{Results of the stacking analysis.}
\begin{tabular}{r c c c c c c c}
\hline
 (1) & (2) & (3) & (4) & (5) & (6) & (7) & (8)   \\
 Class & \# sources & mean z & mean $P_{\rm r(1.4GHz)}$ & $f_{100\mu m}$ & $f_{160\mu m}$ & $S/N_{100\mu m}$ & $S/N_{160\mu m}$ \\
       &            &        & [log (W Hz$^-1$)]& mJy & mJy & &  \\
\hline
SFG & 21 & 0.86 & 23.12 & 7.78 & 12.90 & 2.9 & 2.7 \\
SFG & 41 & 1.88 & 23.98 & 2.39 & 8.08 & 3.6 & 2.9 \\
RQ AGN & 12 & 0.68 & 22.90 & 1.49 & 6.48 & 1.7 & 4.2 \\
RQ AGN & 24 & 2.03 & 24.04 & 4.88 & 11.10 & 2.9 & 2.4 \\
RL AGN & 49 & 0.57 & 22.90 & 1.66 & 3.04 & 2.6 & 1.9 \\
RL AGN & 37 & 1.25 & 23.98 & 1.09 & 2.85 & 1.3 & 1.4 \\
\hline
\end{tabular}
\label{tab_stacking}
\end{table*}

To compute the average FIR luminosity for each sub-sample of sources,
we build their average optical-to-FIR SED and then we proceed as
described in Sec. \ref{sec_Lfir}. Stacked flux densities below the
2$\sigma$ threshold are treated as upper limits in the fitting
procedure.

The corresponding average $SFR_{\rm FIR}$ are plotted in
Fig.\ref{fig_sfr_comparison} as large full symbols, with colors slightly lighter than the ones of the 
their corresponding source class. The average $SFRs_{\rm FIR}$ of
PACS-undetected SFGs and RQ AGNs lie within the scatter of the
corresponding linear relations. We can therefore conclude that the two
SFR tracers are equivalent for our populations of RQ AGNs and SFGs
regardless of their IR brightness.

The average $SFR_{\rm FIR}$ of the RL AGNs not detected by
\textit{Herschel} (the majority of the population) has been derived taking into account the upper limits obtained with the stacking technique in the PACS bands.
Those average values are significantly smaller than the average $SFR_{\rm r}$ obtained using the mean radio power in each bin. This confirms that their radio
luminosity has a significant contribution from the jets and therefore cannot be used to estimate their star-formation rate.
We also note that, in both low and high power sources, RL AGNs have $SFR_{\rm FIR}$ lower compared to the other two classes of sources meaning that the host galaxies of RL AGNs are on average more passive.

\section{SFR versus stellar mass}
\label{sec_SFR_vs_Mstar}
The SF activity of a galaxy can occur in two different modes \citep[e.g.,][]{daddi10a,genzel10}: a starburst one, probably
triggered by major mergers or in dense SF regions; and a quiescent
one, associated with secular processes, which is
observed in the majority of the SFGs. In this second mode, the
SFR is correlated with the stellar mass of the galaxy, forming the so called
``main sequence'' (MS) of SFGs. The MS has nearly the same slope both at low and high redshift but
the normalization increases of about a factor of 20 from the local Universe to $z\approx2$
\citep[e.g.,][]{noeske07,daddi07a,elbaz07,pannella09,peng10,rodighiero11,wuyts11b}.

\begin{figure*}
	\begin{tabular}{c c}
		\centering
	\includegraphics[width=\columnwidth]{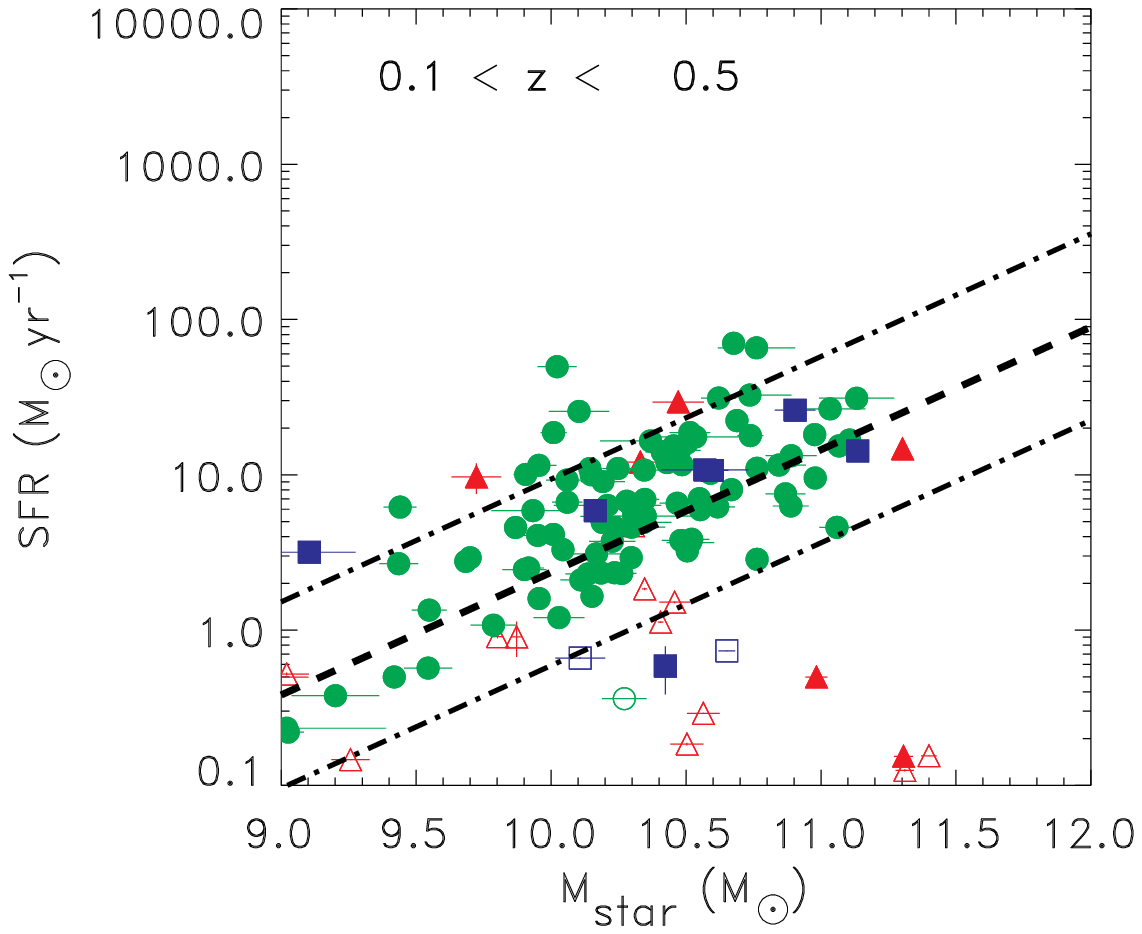} &
	\includegraphics[width=\columnwidth]{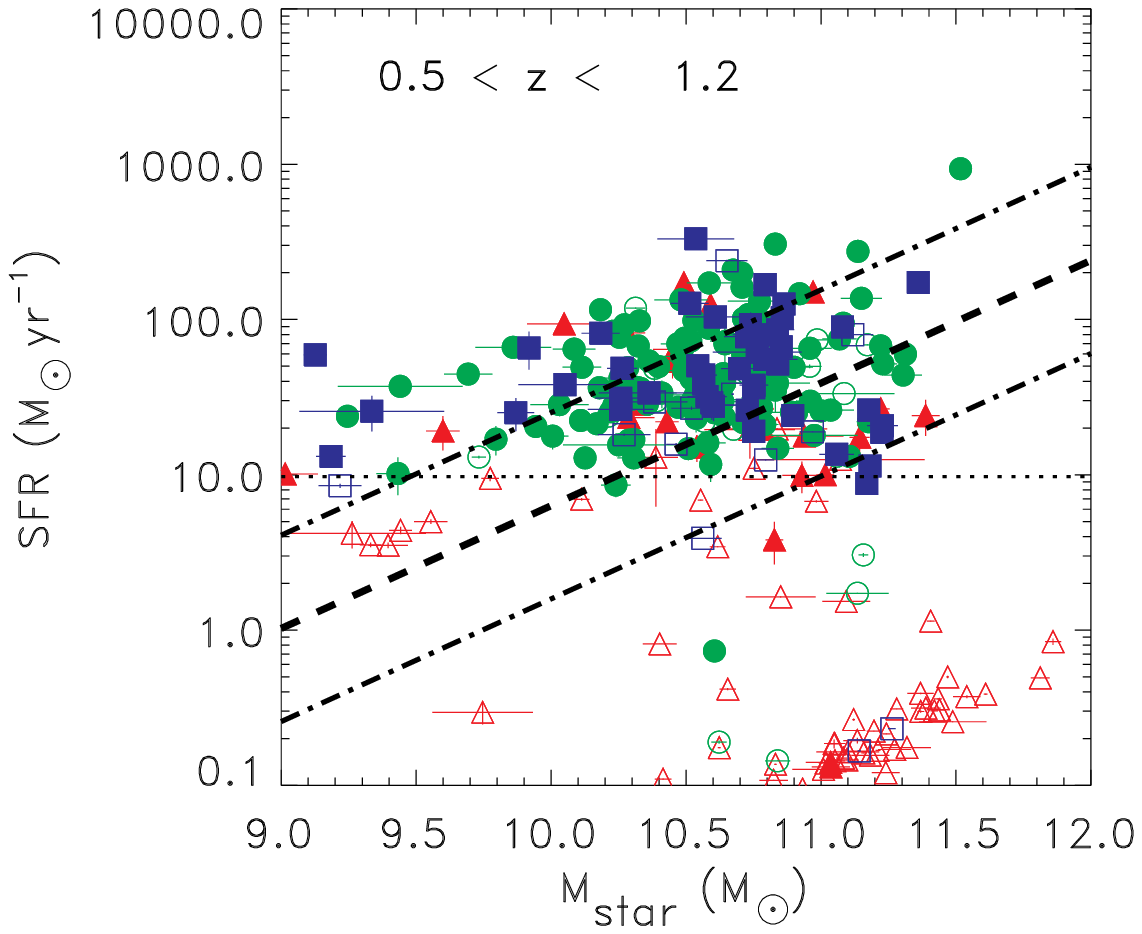}\\
	\includegraphics[width=\columnwidth]{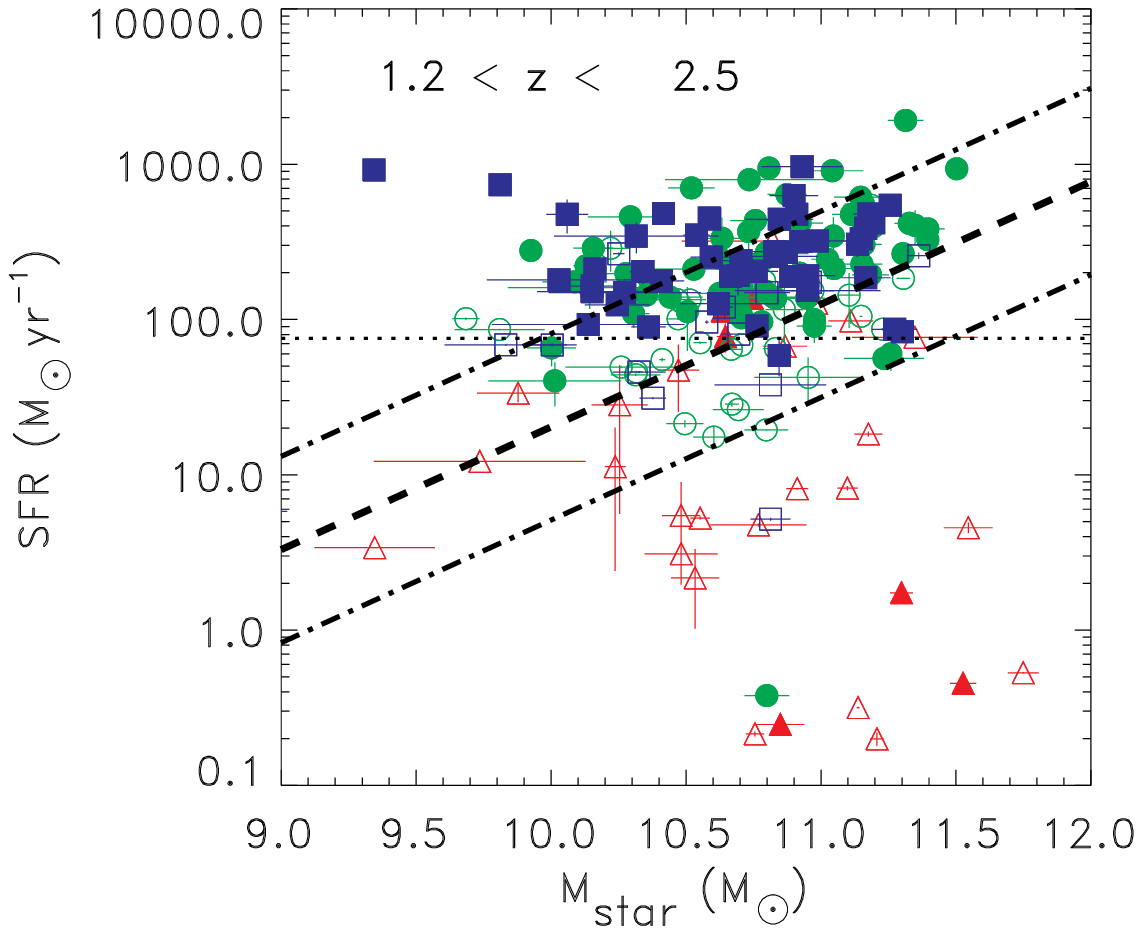} &
	\includegraphics[width=\columnwidth]{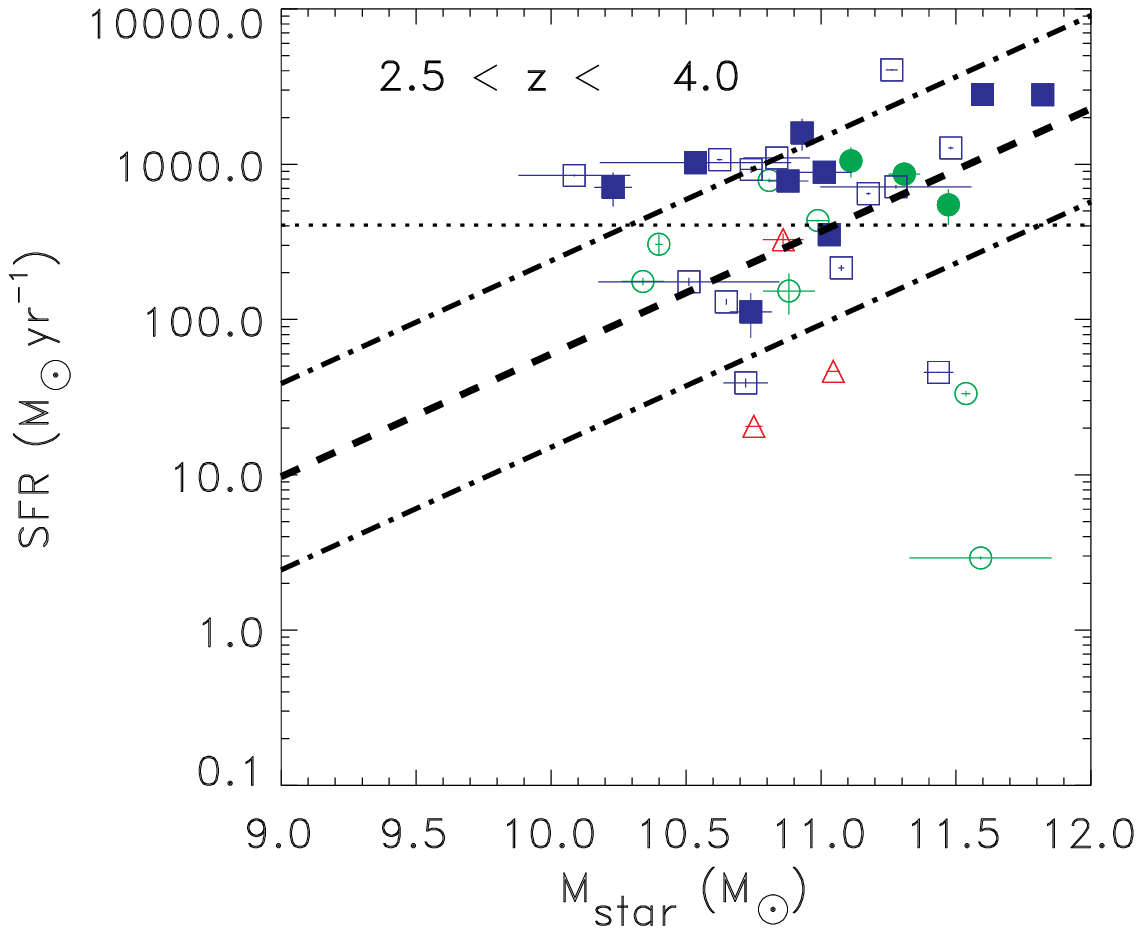}\\

	\end{tabular}
 \caption[SFR versus stellar mass for different redshift bins.]{\small{SFR versus stellar mass for different redshift bins. The symbols are the same as in Figure \ref{fig_sfr_comparison}. The horizontal dotted lines marks the minimum SFR accessible to our survey in each redshift bin (see Text for details). The dashed lines indicates the position of the MS at the average redshift of the sources in each panel. The loci of starburst and passive galaxies are delimited by the dot-dashed lines ($\pm$0.6 dex) above and below the MS respectively.
}}
 \label{fig_SFR-Mstar}
\end{figure*}

In this framework, it is then interesting to locate our sources on the
SFR-stellar mass plane. In Figure \ref{fig_SFR-Mstar}, we plot the
$SFR_{\rm FIR}$ as a function of the galaxy stellar mass for our
objects divided in four redshift bins.  Having already verified
that the radio and the FIR luminosities are equivalently good tracers of the SFR for
our SFGs and RQ AGNs, we decided to use in the following the SFR
derived from the FIR luminosity because it allows us to carry on the
analysis also for RL AGNs for which the $SFR_{\rm r}$ would instead be
meaningless. Moreover, since the $SFR_{\rm FIR}$ is most widely used
in the literature, it makes easier the comparisons with previous
studies. Stellar masses are computed using a two component
(AGN+galaxy) optical-to-MIR SED fitting technique as described in
\citet{bonzini13}, assuming a Chabrier IMF.

The horizontal dotted line in each panel\footnote{In the first redshift bin the minimum SFR is 0.07 $M_{\odot}$ yr $^{-1}$} of Fig. \ref{fig_SFR-Mstar} marks the minimum SFR
($SFR_{min}$) accessible to our survey, i.e. the SFR that correspond
to the flux density limit of the VLA observations at the minimum redshift
in the specified bin, based on equation \ref{eq_sfr_radio} and
corrected for a Chabrier IMF. Note that SFGs and RQ AGNs mostly lie
above these lines, being radio emission in both classes associated
to star formation related process. RL AGN instead are detected even
when they have SFR well below the $SFR_{min}$ since their radio
emission is boosted by the presence of the jets.

The dashed lines indicates the position of the MS at the average
redshift of the sources in each bin. We adopt the following law for
the redshift evolution of the MS:
\begin{equation}
\log(SFR(M,z))=-7.77+0.79\log(M)+2.8\log(1+z) 
\end{equation}
where $M$ is the stellar mass expressed in unit of solar masses. The slope and the redshift evolution are based on the results of \citet{rodighiero11} and the normalization provides the best agreement with our data (see Appendix \ref{sec_model_param} for details) and it is consistent with the normalizations adopted in the literature (\citet{noeske07} for the local universe; \citet{elbaz07} and \citet{daddi07a} for $z\sim $1 and $z\sim $2, respectively).
The dot-dashed lines above and below the MS correspond to $\pm$0.6 dex. 

Due to our radio flux density limit, we are able to detect MS galaxies
with low stellar mass ($M<10^{10.5}M_{\odot}$) only in the first redshift bin
while at higher redshift we probe the bulk of the main sequence only
at the high mass end.

We note that RQ AGNs occupy the same locus in the $M_{star}-SFR$ plane as SFGs suggesting that the majority of the host galaxies of radio selected RQ AGNs are not significantly different from the inactive galaxies population (see Sec. \ref{sec_AGN_content} for a more detailed discussion).
The fact that the relative fraction of SFGs over the RQ AGN decreases as a function of redshift is due to a combination of selection and evolutionary effects as further discussed in the next session and in \citet{padovani15}.

\section{Specific Star Formation Rate}
\label{sec_sSFR}
Since the SFR is correlated with the stellar mass, a useful quantity
to describe the SF regime of a galaxy is its specific SFR
(sSFR), i.e. the SFR divided by the stellar mass.
%We will refer to sources with particularly high sSFR as "starburst" objects  
Sources with sSFR particularly higher with respect to the MS are undergoing extremely intense
star formation activity, possibly triggered by major mergers, while
passive galaxies are characterized by very low sSFRs. In this work,
we will define as "starburst" those sources whose distance with respect to the MS (i.e., $\Delta \log (sSFR)_{MS} = \log [sSFR(galaxy)/sSFR_{MS}(M_{star},z)]$) is larger than 0.6 dex, and we will call ``passive galaxies'' those with $\Delta \log (sSFR)_{MS} <0.6$ dex (similarly to
\citep{rodighiero11}).

In the following, we will discuss the star-formation properties of the
three classes of sources introduced in Sec. \ref{sec_jets_or_SF}. We
will treat RL AGN separately as their properties are significantly
different from the other two classes and the flux density limit of the
VLA survey has a less direct impact on the minimum star-formation rate
detectable in these objects (see Sec. \ref{sec_SFR_vs_Mstar}).

\begin{figure*}
	\begin{tabular}{c c c c}
		\centering
	\includegraphics[trim=0.1cm 0.5cm 2.2cm 0.4cm, clip=true,width=0.45\columnwidth]{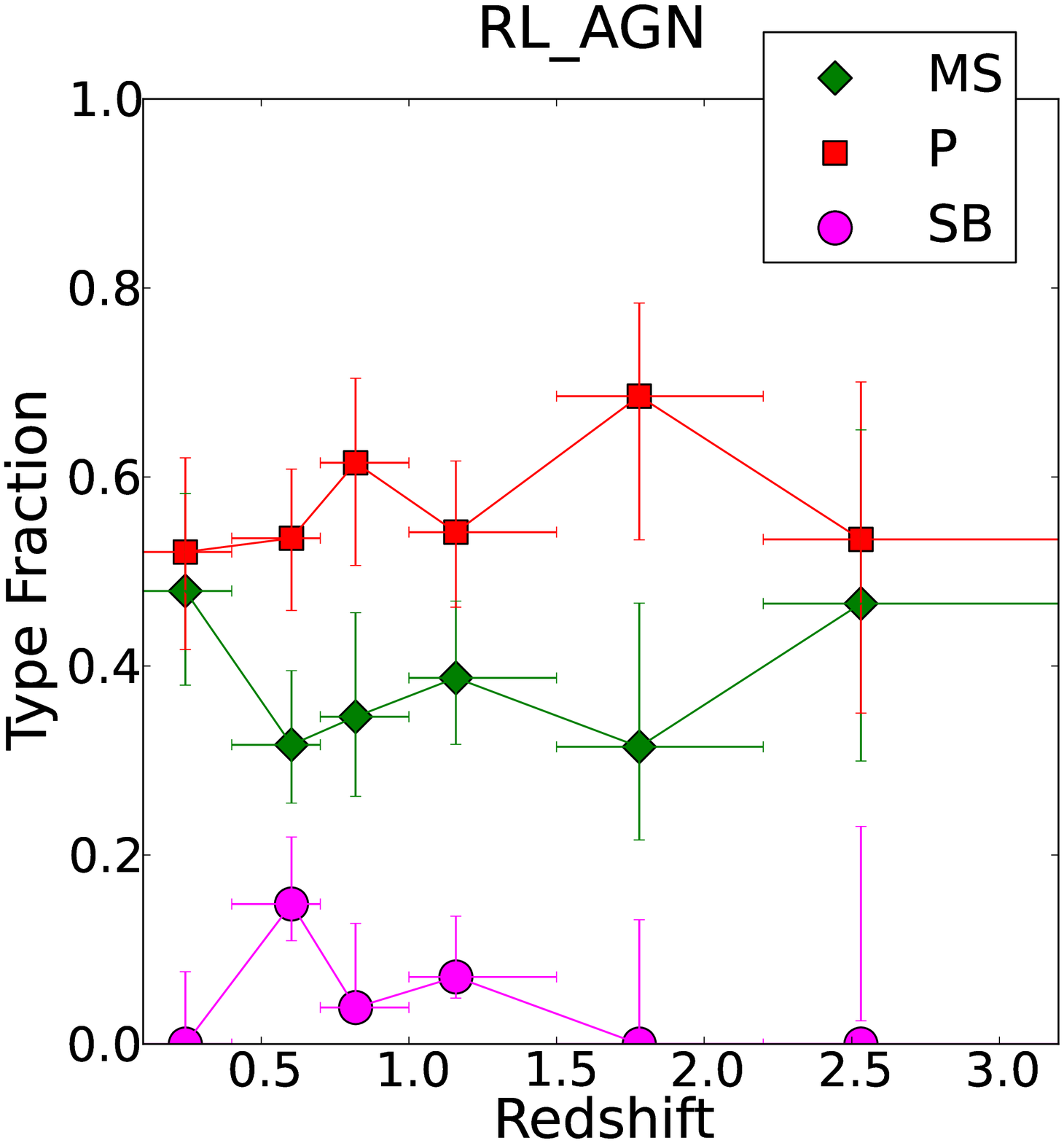}&
	\includegraphics[trim=0.1cm 0.5cm 2.2cm 0.4cm, clip=true,width=0.45\columnwidth]{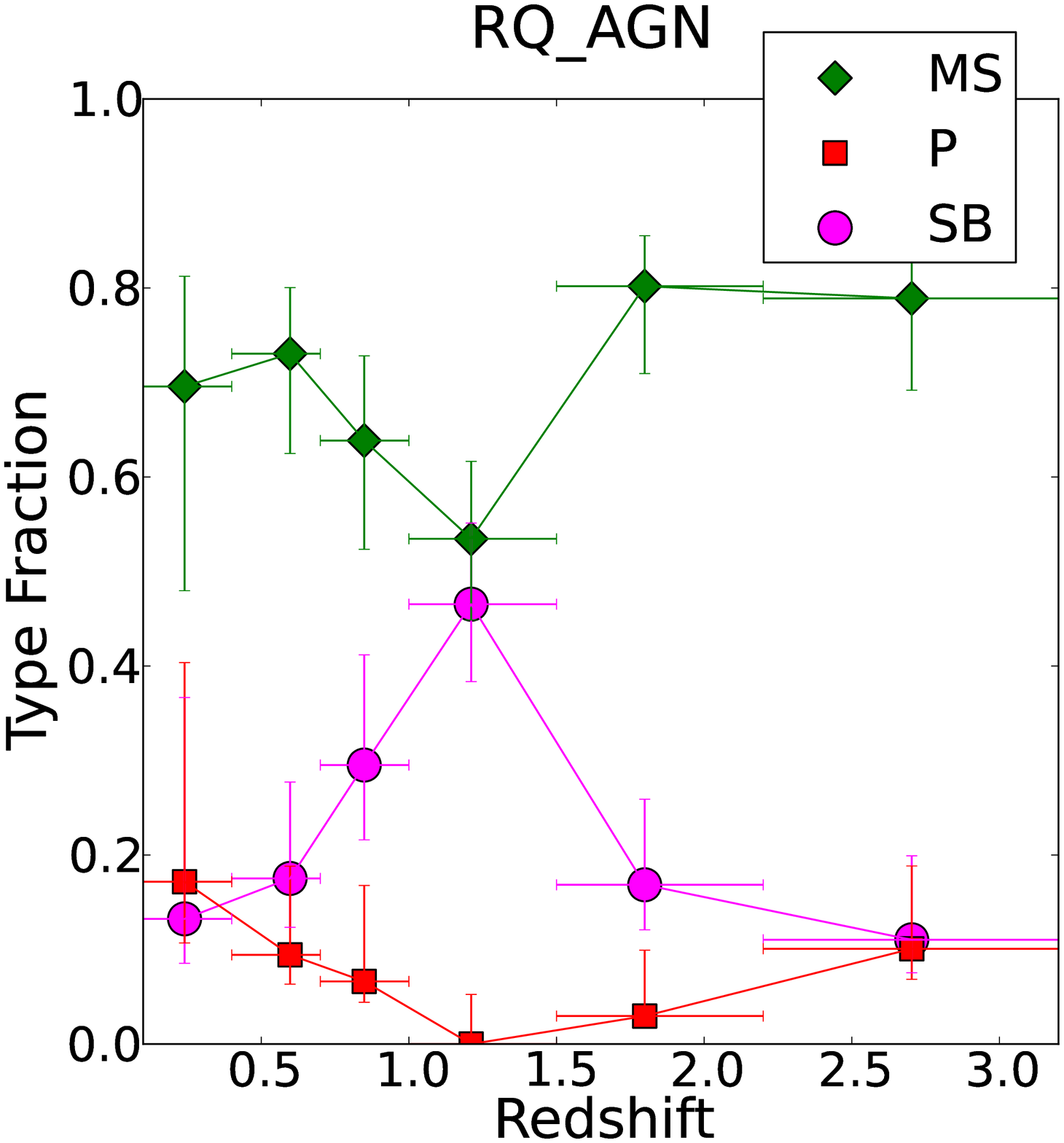} &
	\includegraphics[trim=0.1cm 0.5cm 2.2cm 0.4cm, clip=true,width=0.45\columnwidth]{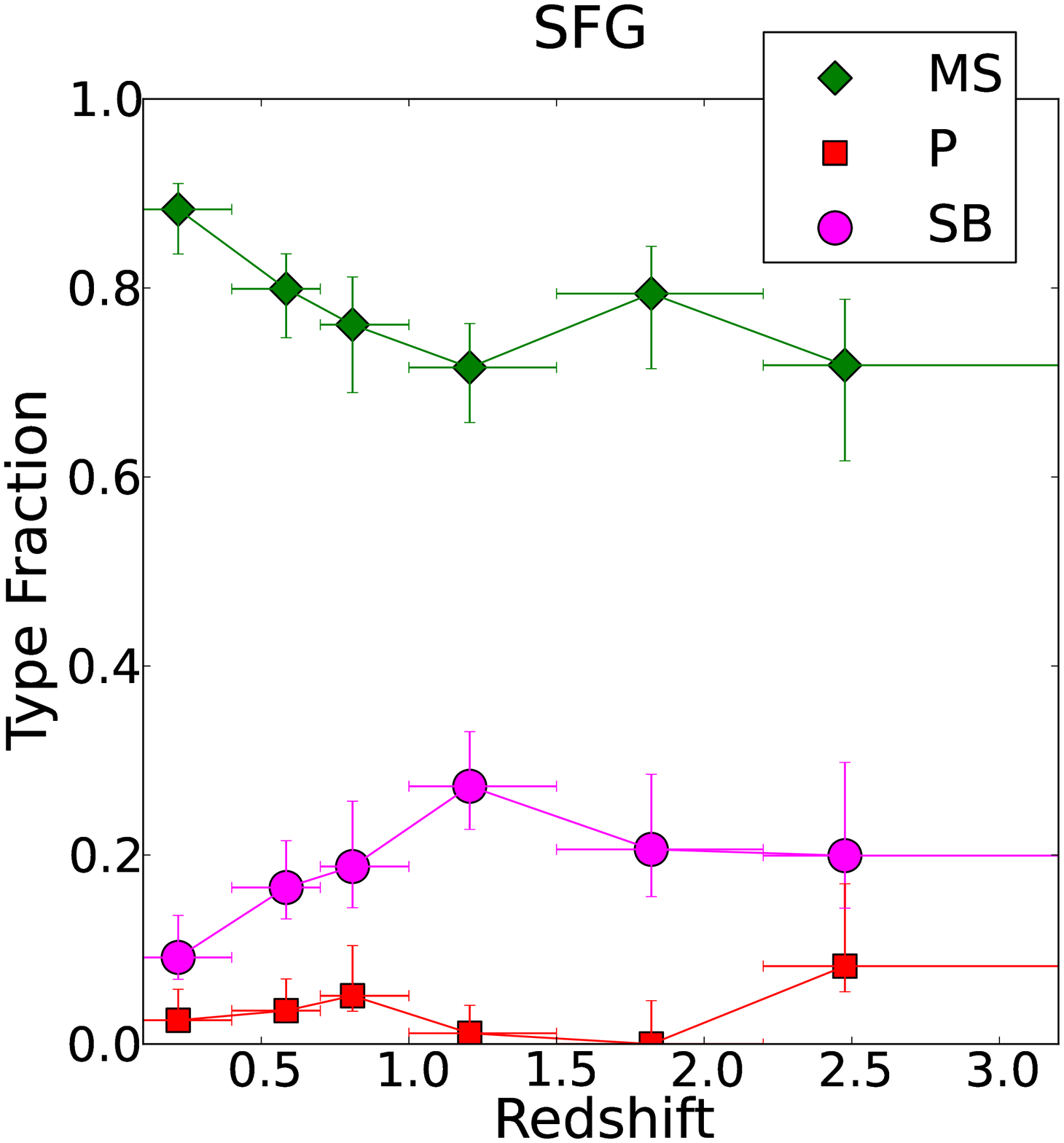} &
	\includegraphics[trim=0.1cm 0.5cm 2.2cm 0.4cm, clip=true,width=0.45\columnwidth]{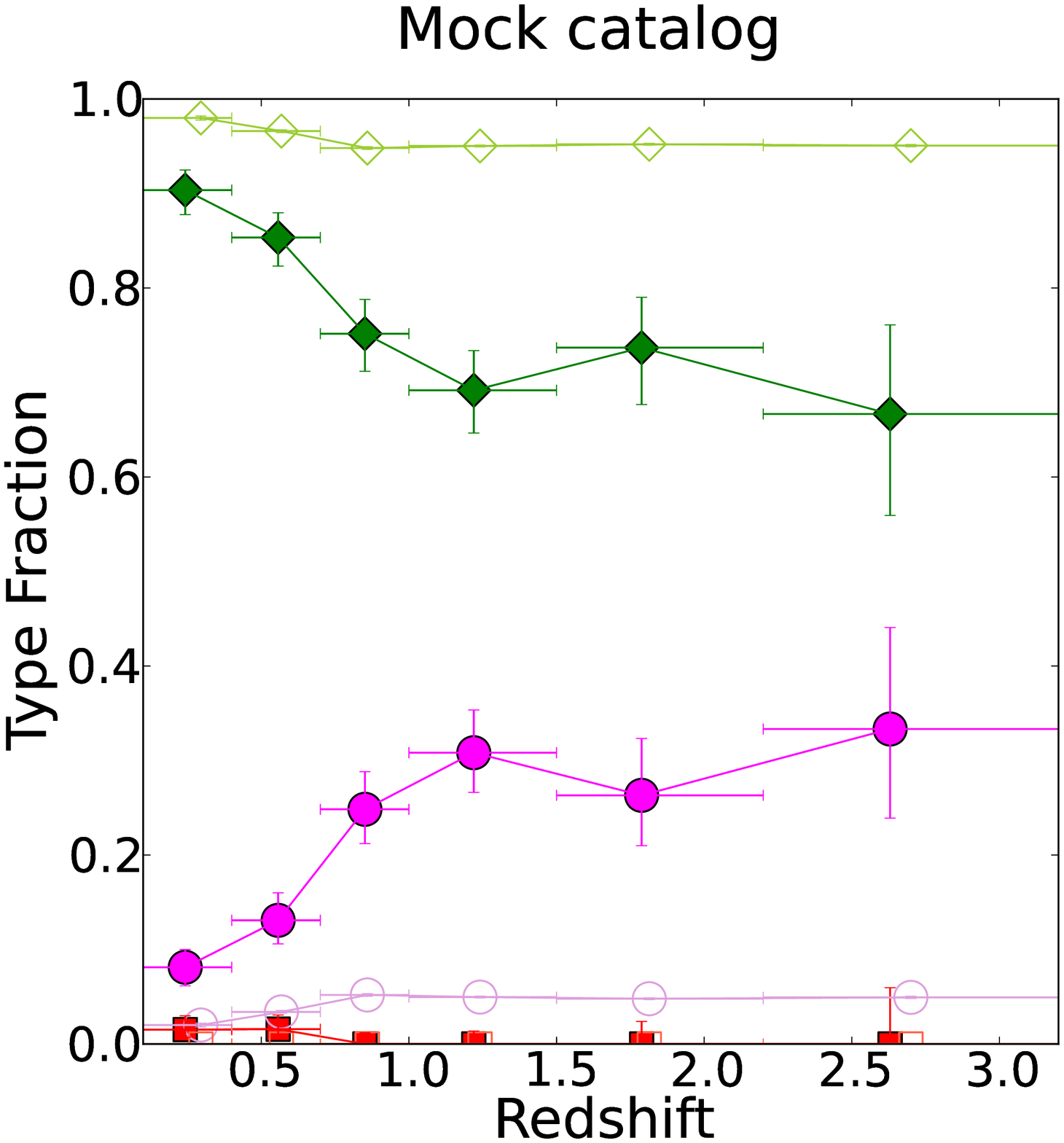} \\

	\end{tabular}
 \caption[Fraction of starbust, MS galaxies, and passive galaxies versus redshift.]{\small{Fraction of starbust (magenta circles), main-sequence galaxies (green diamonds), and passive galaxies (red squares) versus redshift. The panels refer from left to right to RL AGNs, RQ AGNs, SFGs and mock catalogue {of star forming} sources. In this latter panel empty (light colors) symbols refer to the mass selected sample, while full (dark colors) symbols to the mock sample "observed" with the same radio flux density limit of our VLA survey.}}
 \label{fig_types_vs_z}
\end{figure*}

\begin{figure*}
	\begin{tabular}{c c c c}
		\centering
	\includegraphics[trim=0.1cm 0.4cm 2.2cm 0.4cm, clip=true,width=0.45\columnwidth]{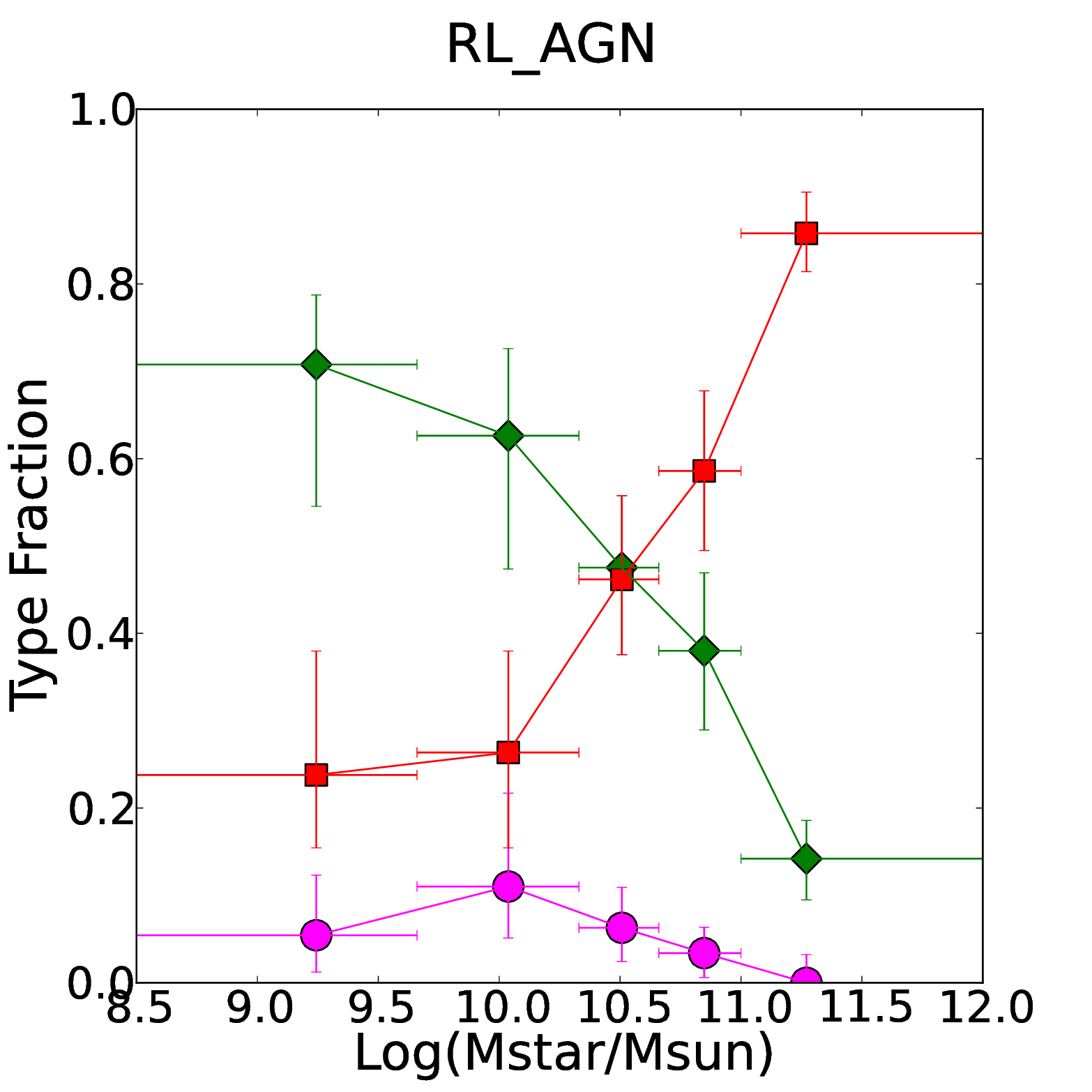}&
	\includegraphics[trim=0.1cm 0.4cm 2.2cm 0.4cm, clip=true,width=0.45\columnwidth]{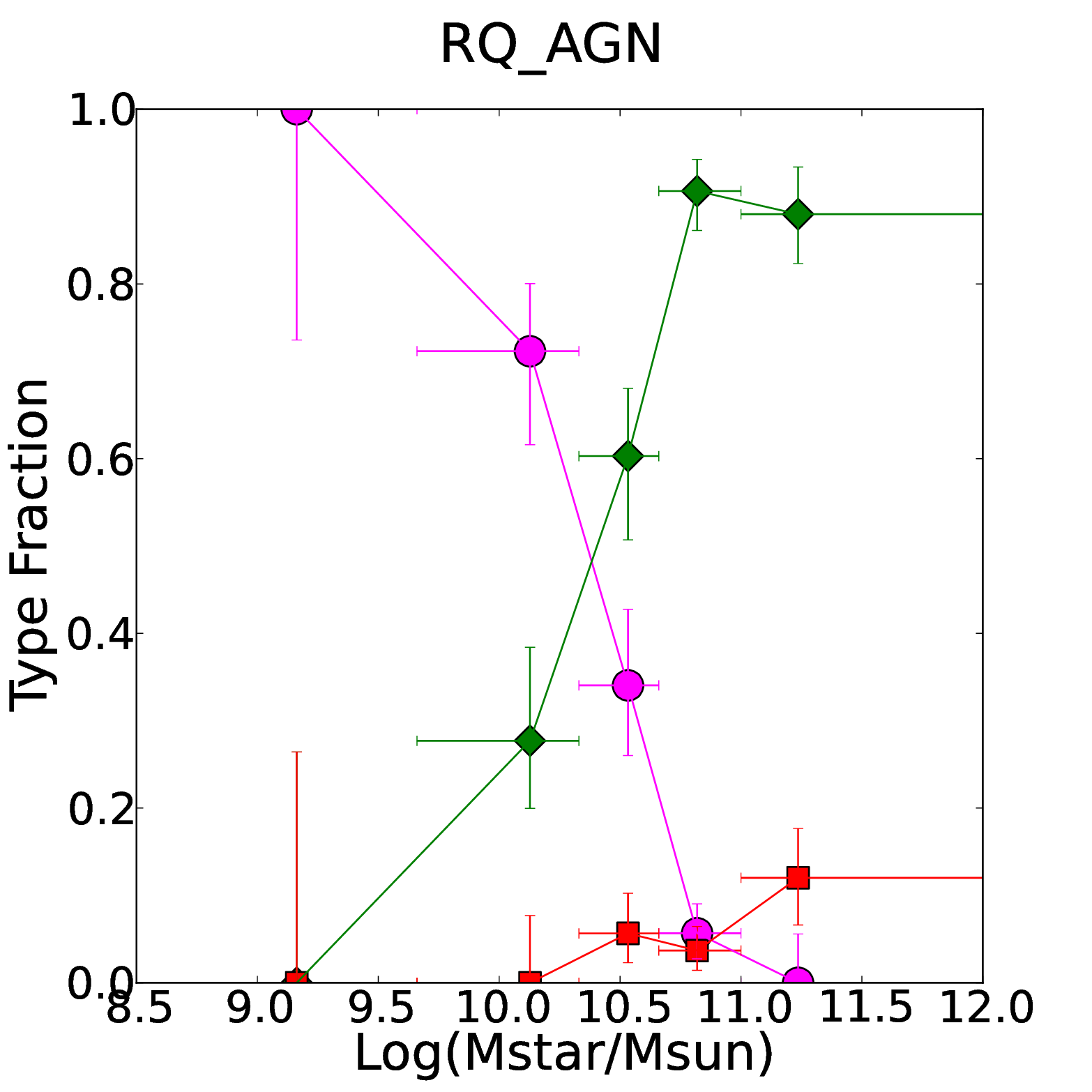} &
	\includegraphics[trim=0.1cm 0.4cm 2.2cm 0.4cm, clip=true,width=0.45\columnwidth]{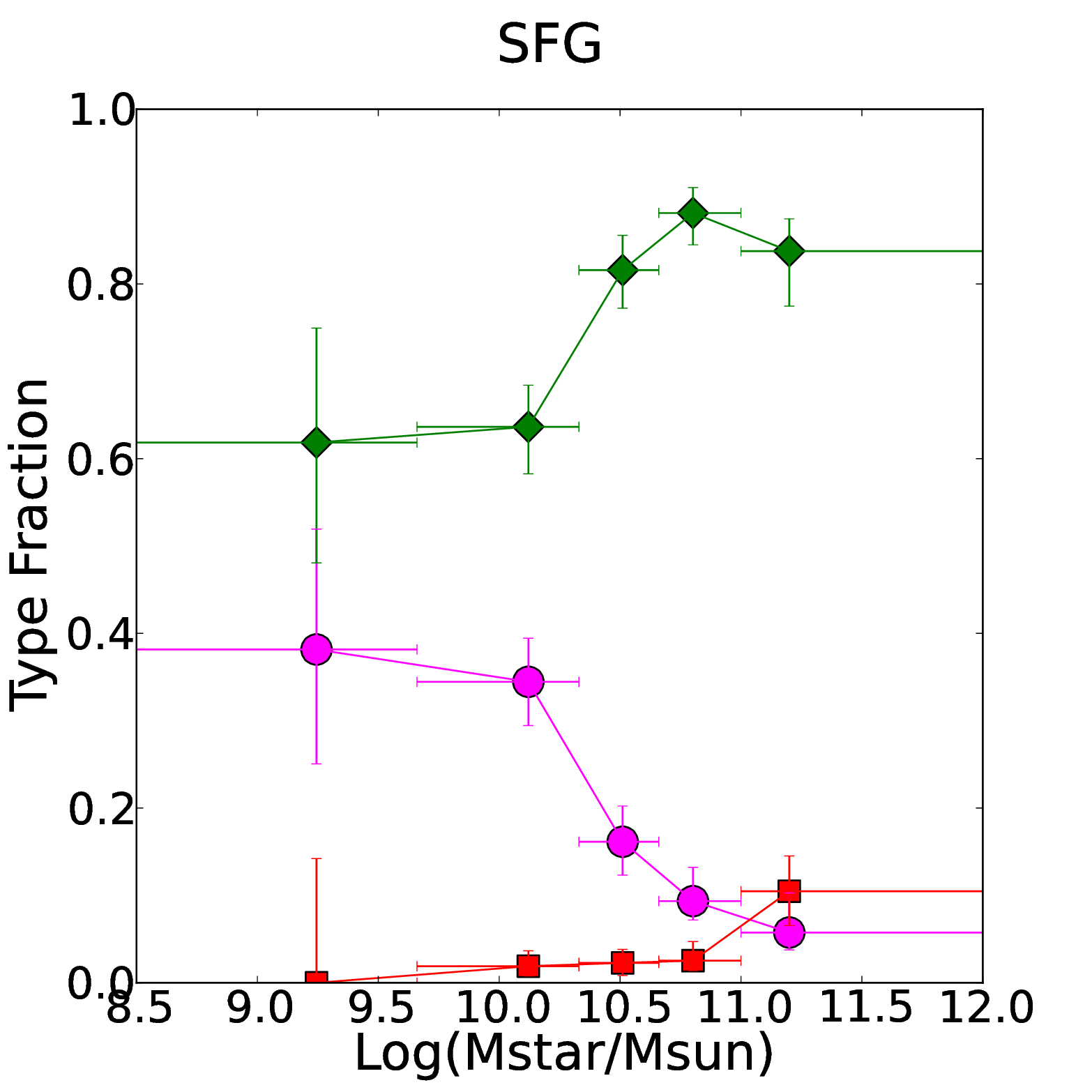} &
	\includegraphics[trim=0.1cm 0.4cm 2.2cm 0.4cm, clip=true,width=0.45\columnwidth]{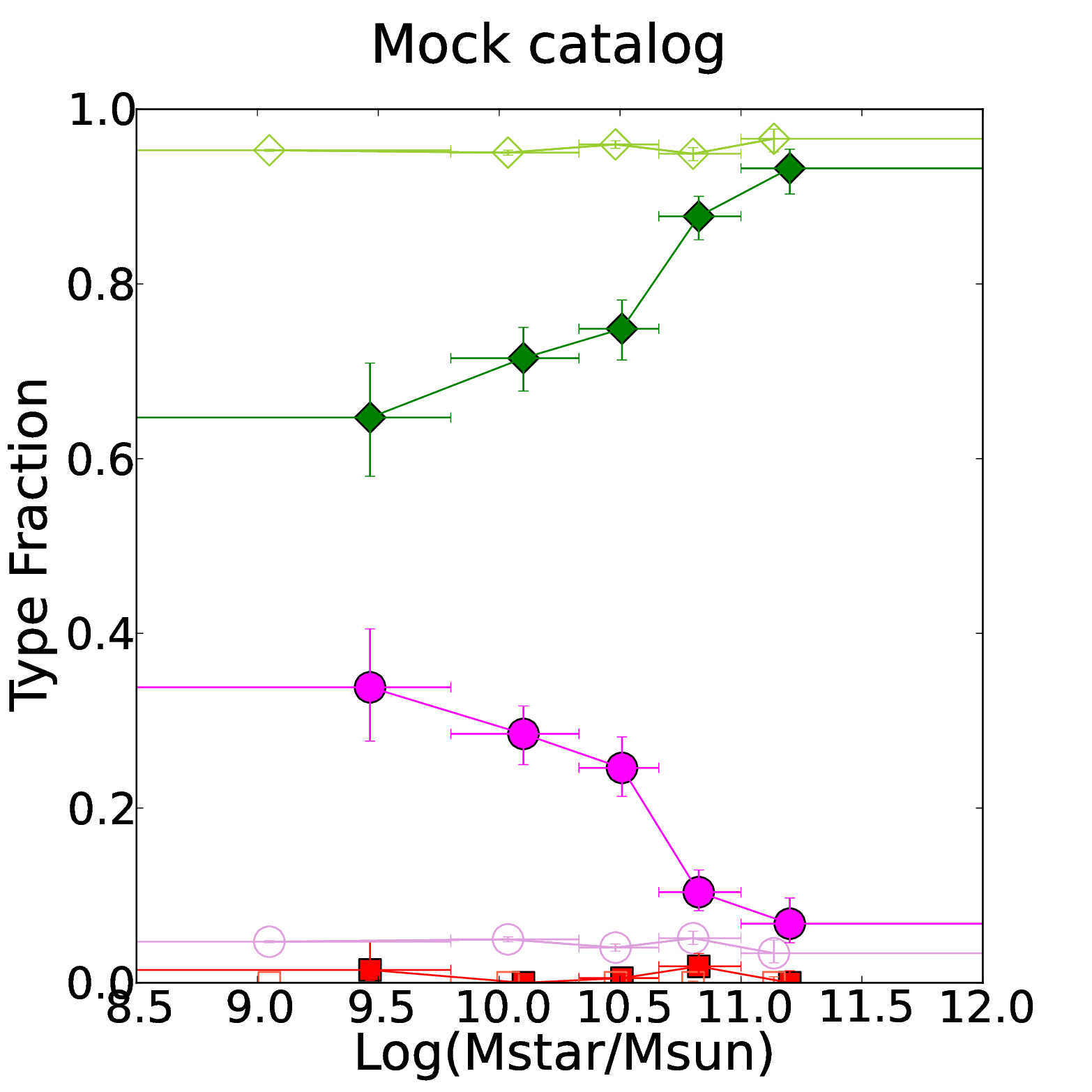} \\

	\end{tabular}
 \caption[Fraction of starbust, MS galaxies, and passive galaxies versus stellar mass.]{\small{Fraction of starbust (magenta), MS galaxies (green), and passive galaxies (red) versus stellar mass. The panels refer from left to right to RL AGNs, RQ AGNs, SFGs and mock catalogue {of star forming} sources. In this latter panel empty (light colors) symbols refer to the mass selected sample, while full (dark colors) symbols to the mock sample "observed" with the same radio flux density limit of our VLA survey.}}
 \label{fig_types_vs_Mstar}
\end{figure*}

\subsection{SF in RL AGNs}
\label{sec_SFinRL}
\begin{figure}
	\centering
	\includegraphics[width=\columnwidth]{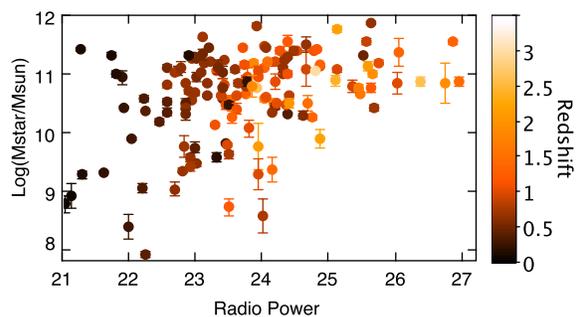}
 \caption[Stellar mass versus radio power for RL AGN.]{\small{Stellar mass versus radio power for RL AGN. The color map shows the redshift of the sources.}}
 \label{fig_Mstar_vs_Pr_RL}
\end{figure}

RL AGNs are generally though to be hosted in massive red and dead
galaxies. Indeed, the majority (56\%) of our RL AGN are hosted in
passive galaxies but a significant fraction of them are in
MS star-forming galaxies and a small fraction (5\%) is even hosted in
starburst systems.  Examples of powerful radio loud AGNs hosted in
starburst galaxies have been found at high redshift (z$>$3)
\citep[e.g.,][]{ivison12}. We have investigated whether the fraction
of actively star forming host galaxies increases with redshift. As shown in the
left panel of Fig. \ref{fig_types_vs_z}, we find no significant trend
up to z$\sim$3.
%The relative contribution of the different types of sources to the RL AGNs population is nearly constant with redshift as shown in the right panel of Fig. \ref{fig_types_vs_z}.
However, we see a strong trend with the stellar mass of the
host (see left panel of Fig \ref{fig_types_vs_Mstar}). Passive
galaxies are the vast majority of RL AGN hosts at masses
$>10^{10.5}M_{\odot}$ while at lower masses they are mostly MS
galaxies.
This trend is similar to what is observed in the overall galaxy
population; the fraction of active galaxies increases as the stellar
mass decreases \citep[e.g.,][]{peng10,ilbert10,ilbert13}. Therefore, 
it seems that there is no strong connection between the presence of
radio jets and the SF activity of the host. On the other hand, we
observe a large spread in stellar masses at low radio power, while the
most powerful RL AGNs are only found in the most massive objects (see
Fig. \ref{fig_Mstar_vs_Pr_RL}) suggesting a link between the maximal
energy that could be released by the jet and the mass of the black
hole, assuming the $M_{BH}-M_{bulge}$ relation. That could also
explain why in shallower radio surveys the fraction of RL AGN hosted
in actively star forming systems is generally lower than what we observe.
As our sources are selected in the radio, it is very unlikely that we are missing bright radio sources with small stellar mass.

Finally, it is also unlikely that we are simply misclassifying SFGs as RL
AGN since these objects are clear outliers from the radio-FIR correlation,
suggesting that there is indeed a significant radio emission
associated with jets from the central black hole.

\subsection{SF in RQ AGNs and SFGs}
\label{sec_SFinRQ}
Once the contribution from RL AGN is removed, a deep radio survey, as the
one presented in this work, can be used to study the SFG population,
its distribution in sSFR, and its evolution with redshift.  Moreover, we
have shown that we can use the radio as an SFR tracer also for the hosts
of most RQ AGNs, including sources in which the AGN emission dominates over the stellar emission in the UV, optical or X-ray bands.
This allows for a more complete census of SFGs and permits to study the AGN contribution in the SFGs population.

Our VLA survey is one of the deepest radio surveys available but still
it is strongly affected by our flux density limit that, as described
in Sec.\ref{sec_SFR_vs_Mstar}, allows us to detect the bulk of the
SFGs population only in the local Universe. Therefore, it is important
to carefully take into account selection effects when drawing any
conclusion. In this context, a comparison with a mass selected catalogue
can be useful.  We used a mass selected mock catalogue \citep{bernhard14} of SFGs based on
the empirical model for the SFG population described in
\citet{sargent12} and \citet{bethermin12}.  We will now briefly describe the model
and the mock catalogue before continuing with the analysis of the SF properties of our radio selected SFGs and RQ AGNs.

\subsubsection{Mock catalogue description}
\label{sec_model_description}
The two basic ingredients to build the mock catalogue are: (i) a
description of the observed SFG mass function (MF) and its redshift
evolution; (ii) a prescription to associate a SFR to each source. For
the MF we adopt a parametrization similar to the one used in
\citet{sargent12}, based on the results of \citet{peng10} and
\citet{ilbert10} (see Appendix \ref{sec_model_param} for more
details). This allows us to build a mass selected SFG catalogue down to
a stellar mass of 10$^{7}$ $M_{\odot}$ (which is smaller than the
minimum mass measured in our sample) for a mock observation on an
area equal to the size of our field (0.285 deg$^2$).

A SFR is then assigned to each source in order to reproduce the
observed MS of SFGs with its redshift
evolution \citep[e.g.][]{daddi07a,elbaz07,rodighiero11,peng10}, and the
distribution of sSFR at a fixed stellar mass. The latter is modelled
as the sum of two Gaussian components \citep{sargent12} one associated
to MS galaxies and the second to starbursts. The full width half maximum (FWHM) of the
two Gaussian, the relative contribution of the two components and its
evolution with redshift are free parameters.  We refer to
\citet{sargent12} and \citet{bethermin12} for a better description of
this model and to Appendix \ref{sec_model_param} for an illustration
of the model parameters adopted in this work.

In order to directly compare this mock catalogue with our radio sample
we need to "observe" it with the same flux density limit of our VLA
survey. To do that, we converted the SFR of the sources in the mock
catalogue to the corresponding radio power according to equation
\ref{eq_sfr_radio}, adding a 0.2 dex random scatter, in agreement
with what we measure (see Sec. \ref{sec_RFC}). We then computed the corresponding radio flux
density given the source redshift and selected only objects with radio
flux densities above 37 $\mu$Jy, the average flux density limit of our
VLA survey.
Finally we added random uncertainties with 0.2 dex dispersion on the
mass and SFR of the mock objects to take into account the
uncertainties affecting real measurements.

Sources in the mock catalogue are flagged as starburst or MS object depending on which of the two Gaussian components they belong to.
Recovering this information in real observations is not possible for
individual sources, but only statistically. Therefore, to compare the
starburst fraction and its dependence on physical parameters
(e.g. stellar mass or radio power) in the mock catalogue and in our
sample, we will adopt the same definition of starburst, MS and passive
based on the distance of the source from the MS as defined in section
\ref{sec_sSFR}.
Note that even if the mock catalogue include in principle only active galaxies, due to the errors we added to the mock catalogues quantities and to the wings of the MS gaussian distribution, some sources in the mock catalogue are "passive" according to our definition.
We also note that, as the RL AGNs are mainly hosted in passive galaxies, the mock catalogue can be used for a comparison with the properties of SFGs and RQ AGNs, where the radio emission traces the SF activity.

\subsubsection{Starburst contribution}
In our observed VLA sample the large majority of SFGs and RQ AGNs are MS objects ($\sim
74\%$). The remaining sources are mainly starbursts ($\sim 22\%$)
with a small contribution from passive galaxies ($\sim 4\%$) (see also Fig. \ref{fig_SFR-Mstar}). The
total observed fraction of SB ($f_{\rm SB,obs}$) is slightly larger in RQ AGNs
compared to SFGs (25\% vs. 20\%). 
We emphasize that these fractions refer to our radio-flux limited sample, and should not be confused with fractions for a mass selected sample. This result echoes what found for a FIR selected sample by \citet{gruppioni13}.

In more detail, we can follow $f_{\rm SB,obs}$ as a function of stellar mass for the two classes of
sources (middle panels in Fig.\ref{fig_types_vs_Mstar}). For both SFGs and RQ AGNs, $f_{\rm SB,obs}$ decreases quickly with increasing stellar mass
as at low masses we are highly biased towards high sSFR due to our
flux density limit. 
The behaviour in RQ AGNs is even more extreme and the SB galaxies become the
majority at masses smaller than $\sim10^{10.5} M_{\odot}$. This is
mainly due to the fact that we are able to detect low mass objects
only at low redshift (see first panel of Fig. \ref{fig_SFR-Mstar}),
but, since the volume that we probe at z$<0.5$ is small, the chance of
detecting AGNs is very low. We are able to sample the MS at low
masses only for more numerous objects as the SFGs.
The effect of the flux density limit becomes clearer looking at the right panel of
Fig. \ref{fig_types_vs_Mstar}; we show the starburst and main sequence galaxies relative fraction for the mass selected
mock catalogue (light colors and empty symbols) and for the same sample after applying the flux density cut (dark colors and full symbols). In the mass selected case, the SB fraction is constant with stellar mass while the "observed" mock catalogue shows the same trend as in the real VLA samples.

More interestingly, we can also follow the evolution of $f_{\rm SB,obs}$, that is equivalent to the relative SB contribution to the SFR
density, as a function of redshift.
Some hydrodynamic simulations suggest an increase of the SB fraction out to z$\sim$2 due to the increase of the merger-induced star formation at high redshift \citep[e.g.,][]{hopkins10}.
In our VLA sample, we observe a fast growth from the local Universe up to z$\sim$1
followed by a flattening both for RQ AGNs and SFGs (see middle panels
of Fig. \ref{fig_types_vs_z}). 
This is not only a consequence of our
flux density limits but also reflects the intrinsic growth of the
starburst contribution from the local Universe to the peak of the star
formation activity. 

Again, the comparison with the mock catalogue allows us to better quantify our results. The evolution of the fraction of starburst in a mass selected sample ($f_{\rm SB,mass}$), is one of the
ingredients of the empirical model used to build the mock catalogue (see Appendix \ref{sec_model_param}) and we can therefore use our observation to constrain the model parameters. %that provide the best match with our observations.
We fixed the local value of $f_{\rm SB,mass}$ to 1.2\% as computed by
\citet{sargent12} as our small field
does not allow us to have enough volume to properly constrain it, while varying the slope of the growth and the redshift above which the evolution stops. 
We find that a growth as $(1+z)^2$ up to z=1 of $f_{\rm SB,mass}$  and constant thereafter is required to match the total $f_{\rm SB,obs}$ in our combined sample of SFGs and RQ AGNs and to reproduce its behaviour as a function of redshift (see Fig. \ref{fig_types_vs_z}).
%A steeper growth of the $f_{\rm SB,mass}$ compared to the one assumed in \citet{bethermin12} has been also confirmed by \citet{gruppioni13}.

In summary, the "observed" mock sample reproduces well the behaviours of the relative fraction of SB, MS and passive galaxies in our SFGs, both as a function of redshift and stellar mass (see the last two panels of Fig. \ref{fig_types_vs_z} and Fig. \ref{fig_types_vs_Mstar}, respectively).  
We note that also when combining SFGs and RQ AGNs, that are the minority in number, the fraction of SB as a function of stellar mass and redshift is consistent with the model within the uncertainties. 
Also the total $f_{\rm SB,obs}$ is the same in the VLA and mock flux limited samples ($\sim20$\%).
Therefore, we conclude that the empirical model of \citet{sargent12} is suitable to describe the star forming galaxy populations with only a little tuning of the model parameters.
 
Therefore, we could use the model prediction to study the behaviour of a mass selected sample rather
than a flux limited ones. In the last panels of both
Fig. \ref{fig_types_vs_z} and \ref{fig_types_vs_Mstar} we show in light colors and empty symbols the
fraction of SB and MS galaxies for the mass selected catalogue.
As expected, the "real" $f_{\rm SB,mass}$ is much lower compared to the
one observed in a flux limited sample and SB sources contribute to
less than 10\% to the total SFR density. This result is also in agreement with the results of \citet{rodighiero11} and the models of \citet{hopkins10}.

\subsection{AGN content as a function of sSFR}
\label{sec_AGN_content}
\begin{figure}
	\centering
	\includegraphics[width=\columnwidth]{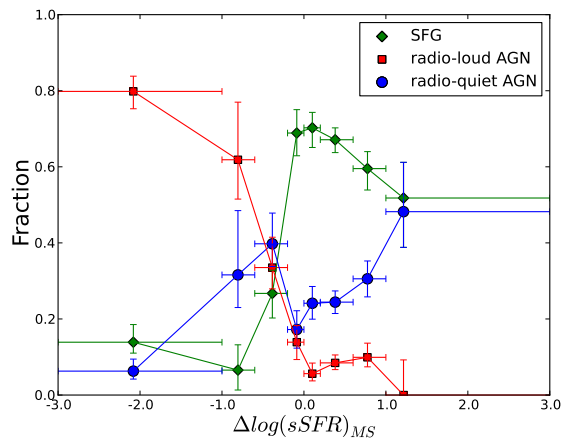}
 \caption{\small{Relative fraction of sources classes as a function of $\Delta log (sSFR)_{MS}$.}}
\label{fig_class_vs_sSFRex}
\end{figure}

In Fig. \ref{fig_class_vs_sSFRex} we plot the relative fraction of the
different classes of sources as a function of the 
distance with respect to the MS (i.e., $\Delta log (sSFR)_{MS} = log [sSFR(galaxy)/sSFR_{MS}(M_{star},z)]$). 
Below the MS (left
side of Fig. \ref{fig_class_vs_sSFRex}) the population is dominated by
RL AGNs. This is due to two main reasons. The first is related to
selection effects since we are sensitive to low sSFR only for sources
whose radio emission is enhanced by AGN activity, i.e., RL AGNs. On the
other hand, the probability of hosting a RL AGN has been observed to be
higher in massive passive galaxies \citep[e.g.,][]{best05a} even if its physical
reason is still unclear.

On the right side of the plot, i.e. above the MS, we observe a rising
of the RQ AGNs fraction from $\sim$25 to 40\%.  To investigate if this
rising can be due to some AGN selection effects we considered two SFG
sub-samples matched in redshift and mass with the RQ AGN
distributions. The trend remains in both cases.  We also checked if it
can be explained by some AGN contribution to the FIR luminosity in two
ways: (i) using the SFR derived from the radio power rather than
$SFR_{\rm FIR}$ and (ii) using the $SFR_{\rm FIR}$ corrected for the
AGN contribution (as describe in Sec.\ref{sec_AGNinFIR}) to compute
the $\Delta log (sSFR)_{MS}$.  Again the fraction of RQ AGNs increases as the $\Delta log (sSFR)_{MS}$
increases. This result is also in agreement with what found by \citet{gruppioni13}, who considered an FIR selected sample and found that starburst sources are dominated by galaxies with AGN-type
SEDs. We will further expand on this result in a coming paper (P. Padovani et al. in prep).
The most extreme SF activity seems therefore often
associated with an active phase of the black hole. Such scenario has
been predicted by simulation of merging systems \citep[e.g.,
][]{dimatteo05, hopkins08}; the instabilities generated by such a
violent event trigger both a burst in the SF and feed efficiently the
central BH.

As discussed in the previous section, sources with SB activity
represent a small fraction of the SFGs population.  Similarly, the
large majority of our RQ AGNs (75\%) are hosted in MS or passive
galaxies.  This fraction only represents a lower limit since our
observation do not fully sample the MS especially at high
redshift. Hence, the most common triggering mechanisms of the black
hole activity is probably related to steady process like gas inflows from the inter galactic medium rather than
extreme events like major mergers \citep[e.g.,][]{mainieri11,rosario12}.

\section{Discussion}
\label{sec_discussion}
We find that radio power is a tracer for the SFR as good as the FIR luminosity not only for SFGs, but also RQ AGNs. Hence, our finding suggests that 
radio emission in RQ AGNs is mainly due to SF. 
As already mentioned in Section \ref{sec_radio_in_RQ}, even if some level of correlation between the radio and FIR luminosity was expected due to the fact that RQ AGNs lie in the same locus as SFGs in the  $q_{24obs}$-redshift plane, the small scatter we find, of only 0.23 dex, was not foreseeable. Indeed, in the MIR, that we use to separate the RL AGNs (see section \ref{sec_jets_or_SF}), the AGN contribution can be strong leading to a scatter in $q_{24obs}$ values for RQ AGNs two times larger compared to SFGs.
Nevertheless, we have shown that
using only two bands, the radio (1.4 GHz) and the MIR band (24
$\mu$m), we can efficiently and with few outliers separate sources powered by the two
different radio emission mechanisms, namely jets and SF. 
 Since the UV-optical and also
the MIR emission can be heavily contaminated by the AGN emission in
powerful AGN, like most of our RQ objects, the radio power could
provide a better estimate of the SF in their host galaxy compared to e.g., UV-based tracers. This result can be particularly useful especially for deep radio surveys without enough FIR coverage.
In addition,
being radio frequencies almost unaffected by dust extinction, it
is suitable for both type I and type II AGNs.

Hypothetically, the good agreement between $SFR_{\rm r}$ and $SFR_{\rm fir}$ for RQ AGNs could be the result of a conspiracy where the contribution to the radio luminosity from tiny jets is exactly compensated by an extra contribution in the FIR. This will boost both the SFR estimates by the same amounts. 
With the current resolution of the VLA data we
cannot exclude the presence of jets in the center of these AGNs
\citep[e.g.,][]{giroletti09} and higher resolution radio observations
with the VLBI would be needed to spatially resolve them.

\subsection{Comparison with a FIR selected sample}

We want to investigate the possibility of using deep radio surveys as a
useful tool to study the cosmic SF history, alternative or
complementary to FIR surveys. We have therefore compared our sample
with the \textit{Herschel}/PACS detected sample.

As described in section \ref{sec_herschel_data} only $\sim$60\% of our
VLA sample has a counterpart in the \textit{Herschel} catalogue of the
E-CDFS.  Excluding the RL AGNs, for which we expected a low detection
rate as most of them are hosted in passive galaxies (but see
Sec. \ref{sec_SFinRL}), the fraction increases to $\sim70\%$. As for
the $30\%$ of unmatched radio sources, this is at least partly due to
the dispersion of the RFC: a radio and a FIR flux density limited
sample would tend to be biased towards sources on opposite sides of
the relation.  To test this hypothesis, we made use of the mass
selected mock catalogue described in Sec. \ref{sec_model_description}.
%To estimate the impact of the different flux density limit,
We "observed" it with a FIR-luminosity limit equivalent to our
37$\mu$Jy flux density limit computing the minimum SFR corresponding
to the radio flux density limit at each source redshift and selecting
only the objects with SFR above this threshold. The resulting
FIR-selected catalogue contains $\sim$780 sources, roughly the same
number of objects as in the radio selected mock catalogue described in
Section \ref{sec_model_description}.
\begin{figure}
	\centering
	\includegraphics[width=\columnwidth]{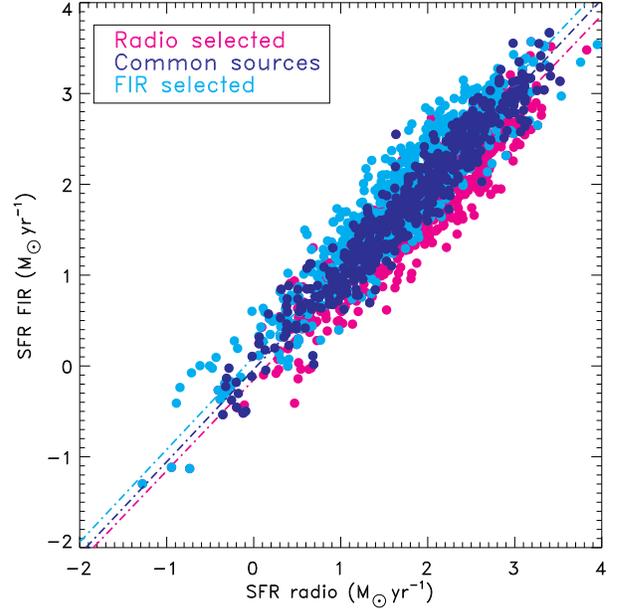}
 \caption[SFR comparison for the mock catalogue.]{\small{SFR derived from the FIR luminosity versus the SFR from the radio luminosity for the sources in the FIR selected mock catalogue (cyan), the radio selected mock catalogue (magenta) and the matched sample (blue). The dashed lines are the linear fit to the data.}}
 \label{fig_sfr_selection_comparison}
\end{figure}
In Fig. \ref{fig_sfr_selection_comparison} we show the RFC for the FIR
selected mock catalogue in cyan, the radio selected in magenta, and for the matched
sources, in blue. The latter are only $~$60\% of the total
number of objects in each catalogue, a fraction similar to the one
obtained matching the real VLA and \textit{Herschel} samples.  The
offset from the 1:1 correlation is $\sim 0.15$ dex above and below for
the FIR and radio selected samples, respectively.  This plot explains
also why most of the VLA sources without PACS detection are just below
the correlation found for the VLA sources with \textit{Herschel}
counterpart in Fig. \ref{fig_sfr_comparison}: being radio-detected
only they have a slight radio excess.

On the other hand, about 40\% of the PACS detected sources have a
counterpart in the VLA image. This means that the FIR observations
reach a lower flux density limit, also thanks to the much deeper maps in
the central GOODS field compared to the outskirts (see
Sec. \ref{sec_herschel_data}).  We have compared the $SFR_{\rm FIR}$
computed in \citet{gruppioni13} for the sources in common and we find
very good agreement with our estimates. Also the stellar masses are
consistent; we find a larger scatter but no systematic shifts.
This reflects the fact that the SFR is well constrained when FIR
photometry is available, while on the stellar masses there are larger
uncertainties, especially in AGN host galaxies, depending on the
fitting technique and especially on the model library adopted
\citep[e.g.][]{ilbert10}.

\begin{figure*}
	\begin{tabular}{c c}
		\centering
	\includegraphics[trim=0.2cm 0.6cm 0.2cm 0.6cm, clip=true,width=1.04\columnwidth]{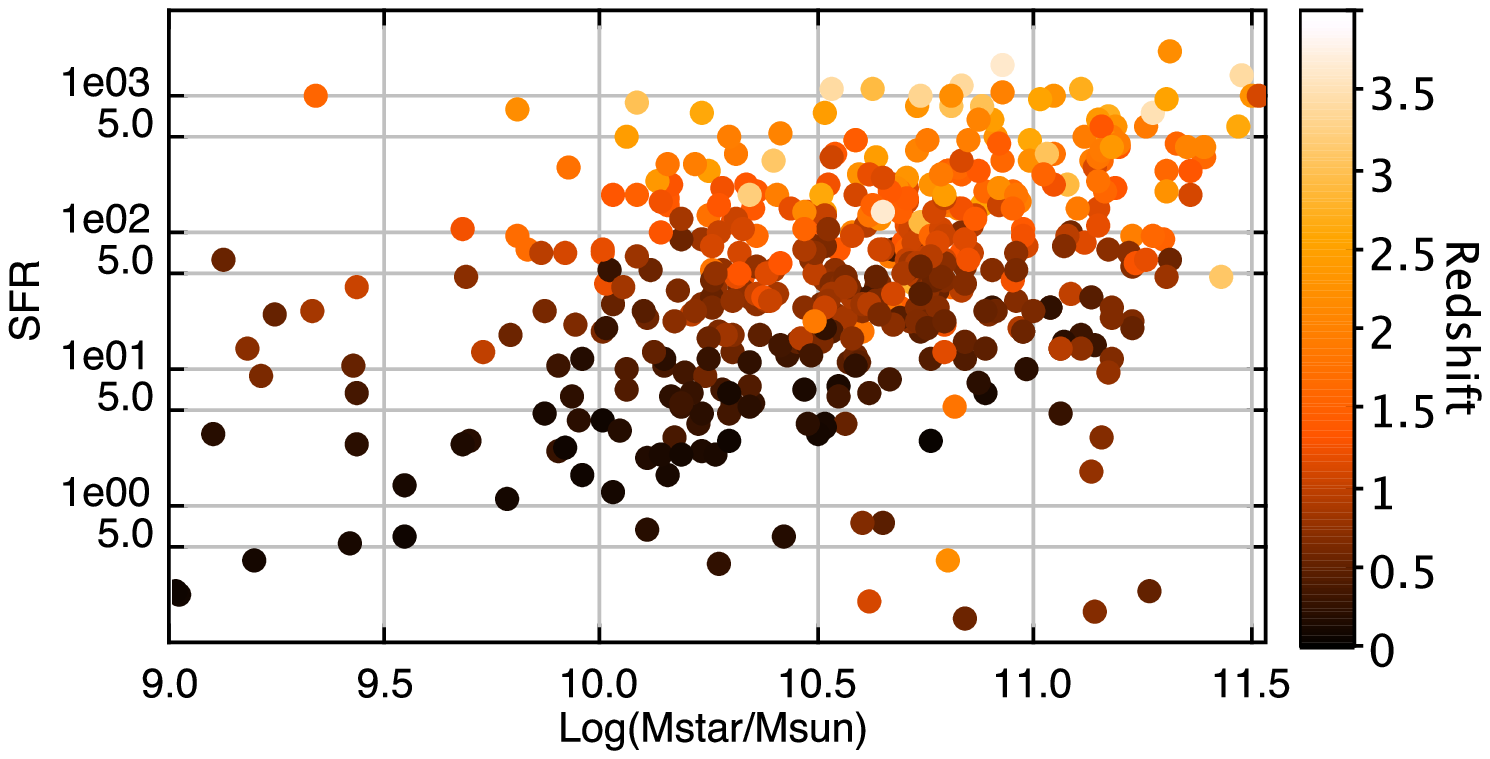} &
	\includegraphics[trim=0.2cm 0.6cm 0.2cm 0.6cm, clip=true,width=1.04\columnwidth]{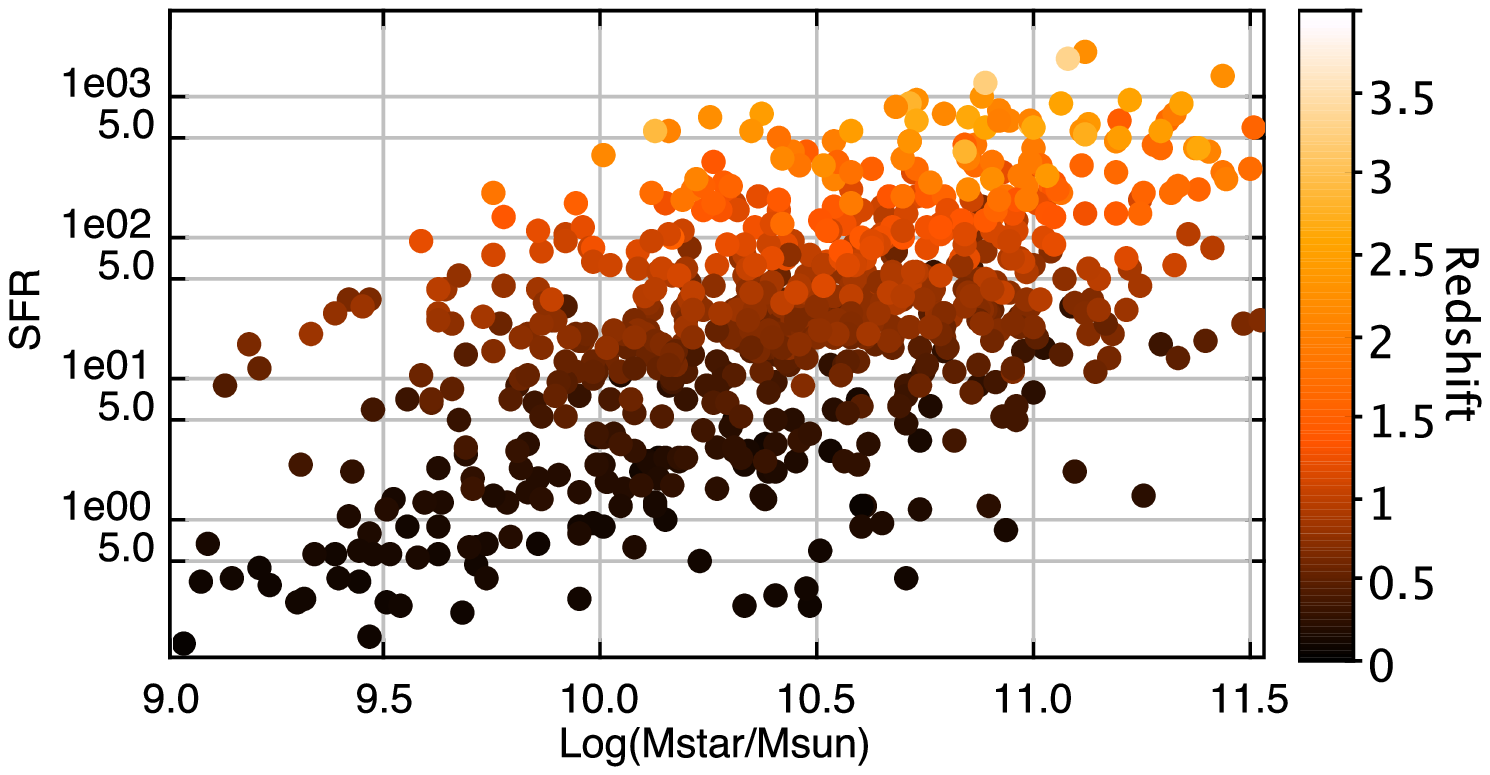} \\
	\end{tabular}
 \caption[SFR versus stellar mass for the VLA and PACS sample]{\small{SFR versus stellar mass for the VLA sample (left) and for the PACS sample (right).}}
 \label{fig_comp_rodighiero}
\end{figure*}

\begin{figure}
	\centering
	\includegraphics[width=\columnwidth]{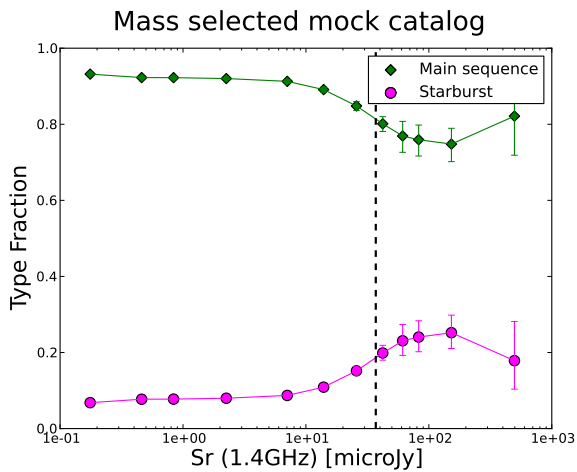}
 \caption[MS and starburst fractions vs. radio flux density.]{\small{Fraction of MS and starburst sources in the mass selected mock catalogue as a function of the radio flux density.}}
 \label{fig_type_vs_Sr_mock}
\end{figure}

In Fig. \ref{fig_comp_rodighiero}, we show the VLA (RQ AGNs and SFGs
only) and the \textit{Herschel} samples on the SFR--stellar mass plane with
the sources coloured according to their redshift.
The two sets of data show the same trends and the same evolution with
redshift. We also find a similar fraction of SB (20\%) in the PACS
sample compared to the radio one.  Being a bit deeper, especially in the GOODS field, the PACS
observations probe better the small stellar
mass end and are able to detect more low SFR sources at low redshift. Already ongoing deeper radio surveys, as well as those planned with future facilities, will allow to probe these small masses
also in the radio \citep[][]{norris13}. Indeed, going down to the
nanoJy sensitivity level, they will be able to probe the bulk of the
SFGs population. An illustration of that is given in
Fig. \ref{fig_type_vs_Sr_mock} where we plot the fractions of MS and
SB galaxies as a function of the radio flux density in $\mu$Jy for the
mass selected mock catalogue. Below about 10 $\mu$Jy the number density
of SB remains nearly constant around 8\%, while at the flux density
limit of our survey (vertical line) we are biased towards higher sSFRs. 

\subsection{What changes using a different SFR tracer?}
As discussed in the previous section, due to the RFC dispersion, a
sample selected with a radio flux density limit would be slightly
biased towards those objects on the right side of the relation,
i.e. with a radio power larger than the one given by the relation for
a given FIR luminosity.

That means that, especially for the sources without \textit{Herschel}
counterpart, the SFR derived from the two different tracers can be
slightly different.  As quantities like the fraction of SB are
extremely sensitive to these changes due to small number statistics, we
investigated how the choice of a different SFR tracer, namely the
radio power, can affect our conclusions. Of course that cannot be
done for RL AGNs as in these objects the radio power is not tracing the
SF activity in the host galaxy but it is highly contaminated by the
jets emission.  For RQ AGNs and SFGs instead we repeated the same
analysis as in Sec. \ref{sec_sSFR} using the $SFR_{\rm r}$ to compute
the $\Delta log (sSFR)_{MS}$ and therefore to determine the level of their SF
activity.

\begin{figure*}
	\begin{tabular}{c c c }
		\centering
	\includegraphics[trim=0.1cm 0.4cm 2.2cm 0.4cm, clip=true,width=0.5\columnwidth]{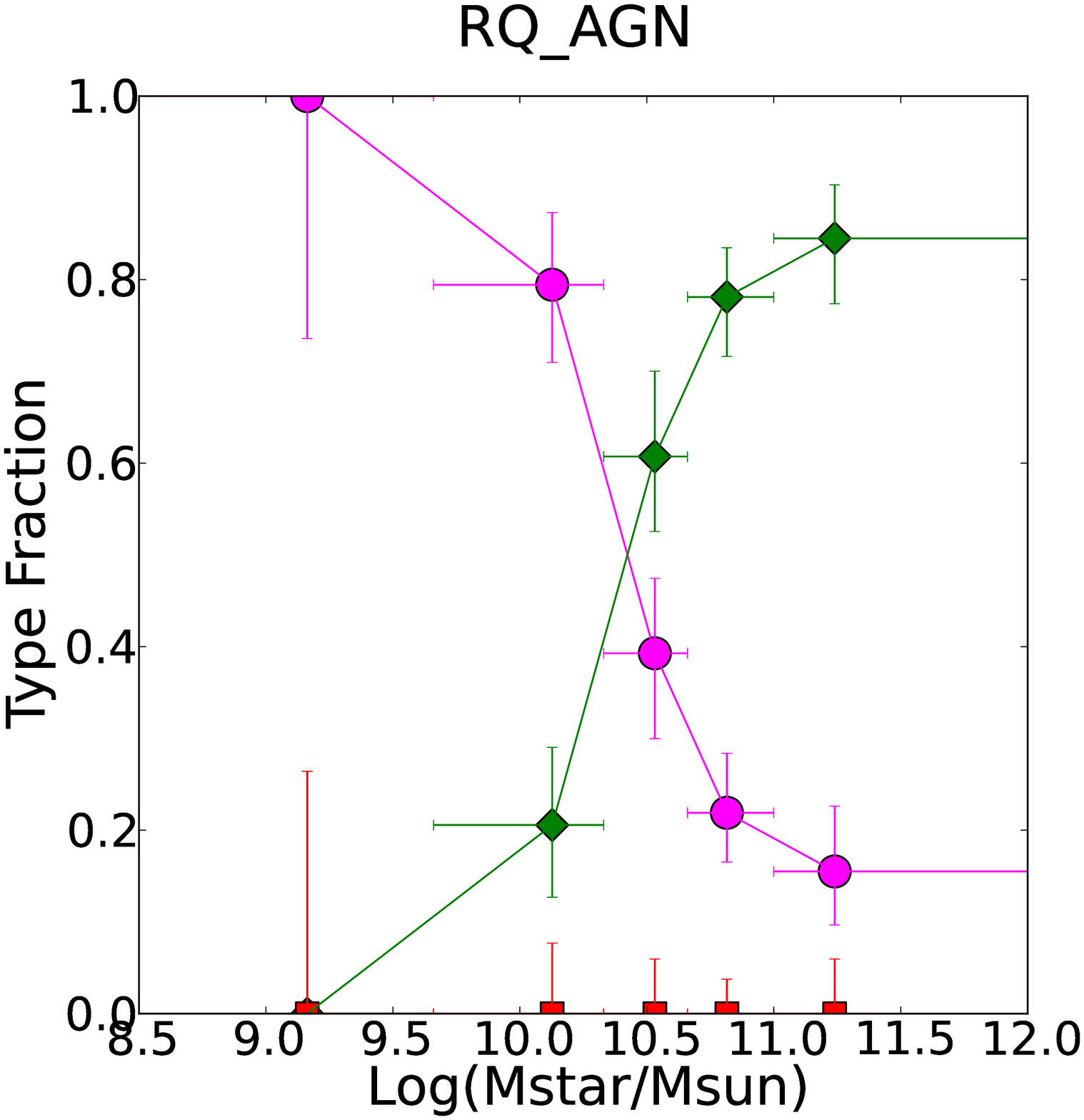} &
	\includegraphics[trim=0.1cm 0.4cm 2.2cm 0.4cm, clip=true,width=0.5\columnwidth]{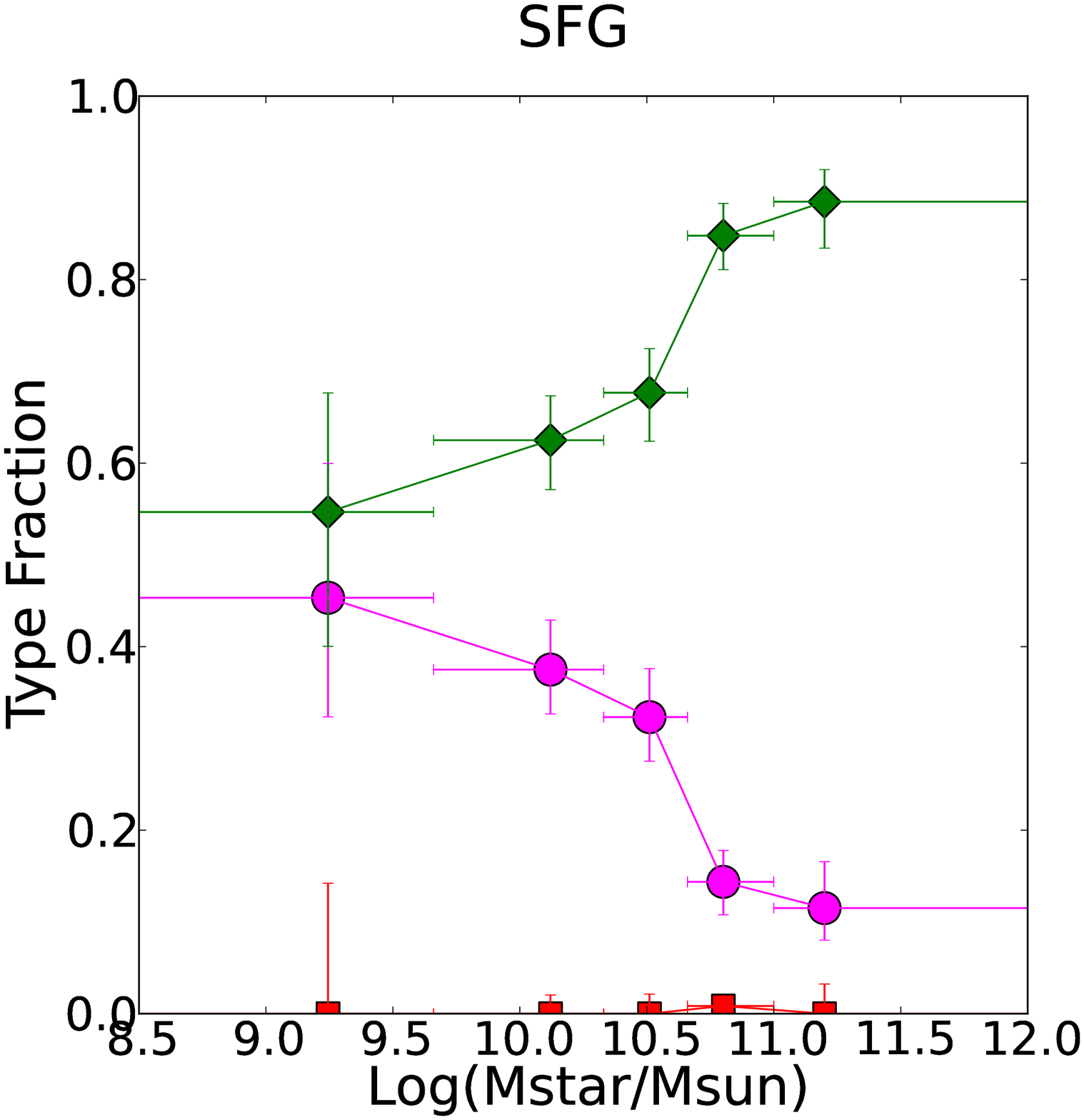} &
	\includegraphics[trim=0.1cm 0.4cm 2.2cm 0.4cm, clip=true,width=0.5\columnwidth]{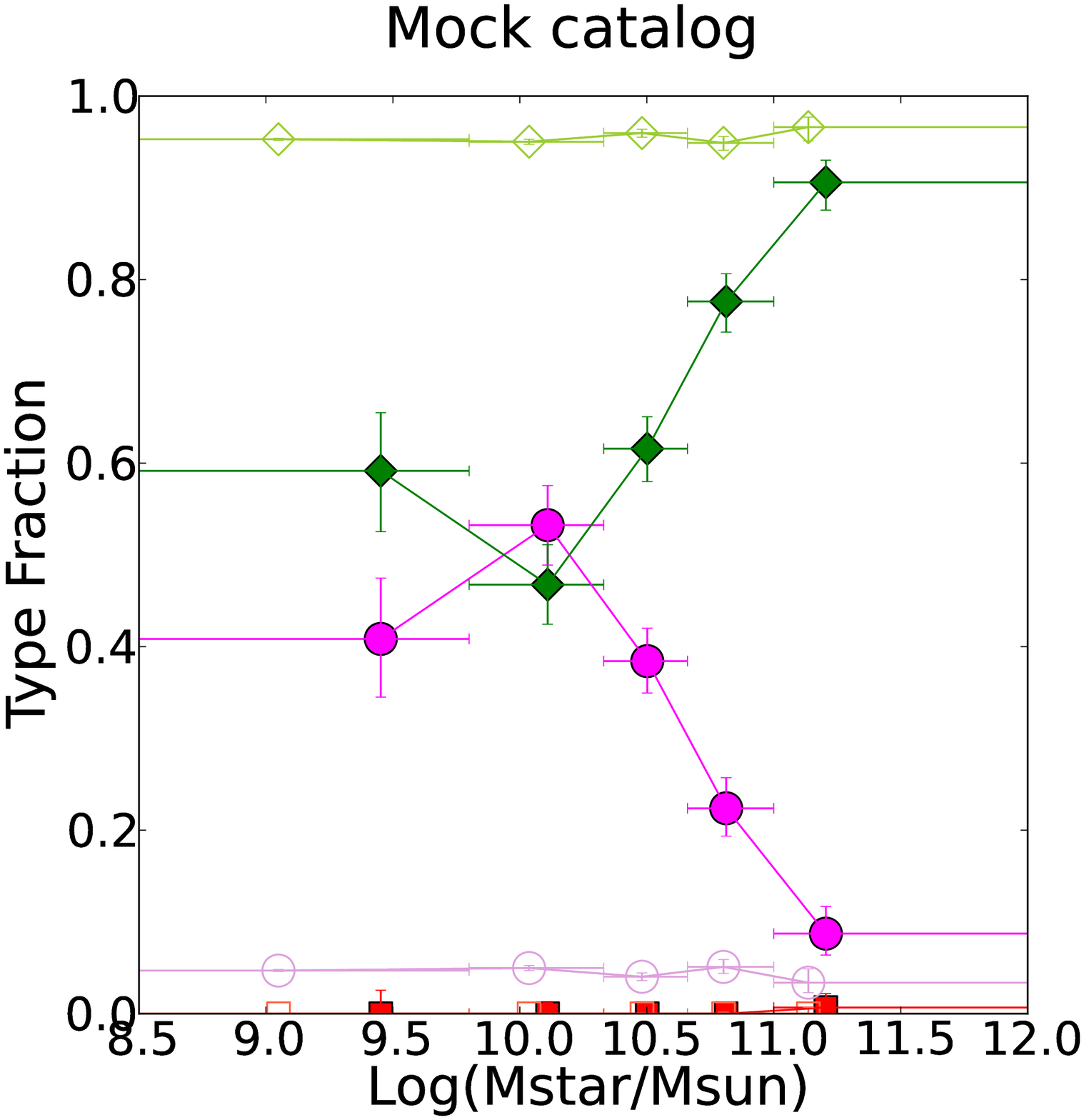} \\

	\end{tabular}
 \caption[Fraction of SB, MS, and passive galaxies derived using the $SFR_{\rm r}$, vs. stellar mass.]{\small{Fraction of starbust (magenta circles), MS galaxies (green diamonds), and passive galaxies (red squares) derived using the $SFR_{\rm r}$, versus stellar mass. The panels refer from left to right to RQ AGNs, SFGs and mock catalogue sources.}}
 \label{fig_types_vs_MstarRadio}
\end{figure*}

The total fraction of SB in this radio flux-limited sample increases to $\sim$30\%\footnote{Note that, in contrast, the fraction of passive goes almost to zero.} but the trends as a
function for example of stellar masses remain the same (see
Fig. \ref{fig_types_vs_MstarRadio}). Also in the radio flux density
limited mock catalogue, if we compute the $\Delta log (sSFR)_{MS}$ from their radio
power we obtain a $f_{\rm SB,obs}$ of 30\%, in agreement with our data.

The larger differences appear at $z>1$, where the
fraction of starbursts is up to a factor of two larger using the
$SFR_{\rm r}$, but this effect is mainly due to the sources without
PACS detection. Indeed, as shown in
Fig. \ref{fig_SFRPACS-24_ratio_vs_z}, the SFR extrapolated from the
optical-to-MIR photometry tends to be underestimated at high
redshift. Hence, the differences between the two SFR tracers appear
larger. As a further consequence, the $f_{\rm SB,obs}$ at high redshift shown
in the central panels of Fig. \ref{fig_types_vs_z} are slightly
underestimated as also suggested by the predictions of the empirical
model.

\subsection{$L_{\rm X}-L_{\rm FIR}$ relation}
The hard band X-ray luminosity ($L_{\rm X}$) is considered another SFR
tracer up to high redshift \citep{norman04}. In starburst
galaxies the  2--10 keV rest-frame emission is dominated by high-mass
X-ray binaries (HMXB).  Being the companion of the accreting object a
short-lived high mass star, the $L_{\rm X}$ is linked to the recent
star formation activity.  At low star formation rates, however, there is 
contamination from low-mass X-ray binaries (LMXB) which do not trace the
immediate  SFR due to their long evolution lifetimes.   The conversion
factor between  $L_{\rm X}$ and the SFR has been calibrated for local
samples \citep[e.g.,][]{ranalli03,persic04,lehmer12}.  It has been
argued that the hard band X-ray luminosity can be used as an unbiased
SFR indicator up  to high redshifts, despite the contamination from
obscured AGN which may be unnoticed in distant, faint sources.

Recently, \citet{vattakunnel12} used the deep X-ray Chandra data
\citep[total integration time of 4 Ms,
  see][]{xue11,lehmer12} and the deep VLA observation
\citep{miller13} to explore the $L_{\rm X}$-$P_r$ relation in
galaxies dominated purely by star formation processes in both bands.
Due to the strict selection process, aimed at discarding any
significant AGN contamination, they obtained a sample of 43 SFGs up to
$z\sim 1.2$. They found a clear linear correlation between radio and
X-ray luminosity in SFG over three orders of
magnitude in this redshift range, consistent with that measured
locally.  They also measured a significant scatter of the order of $\sim 0.4$
dex, higher than that observed at low redshift, implying an intrinsic
scatter component, or some residual AGN contamination. 

\begin{figure*}
	\begin{tabular}{c c }
		\centering
	\includegraphics[width=\columnwidth]{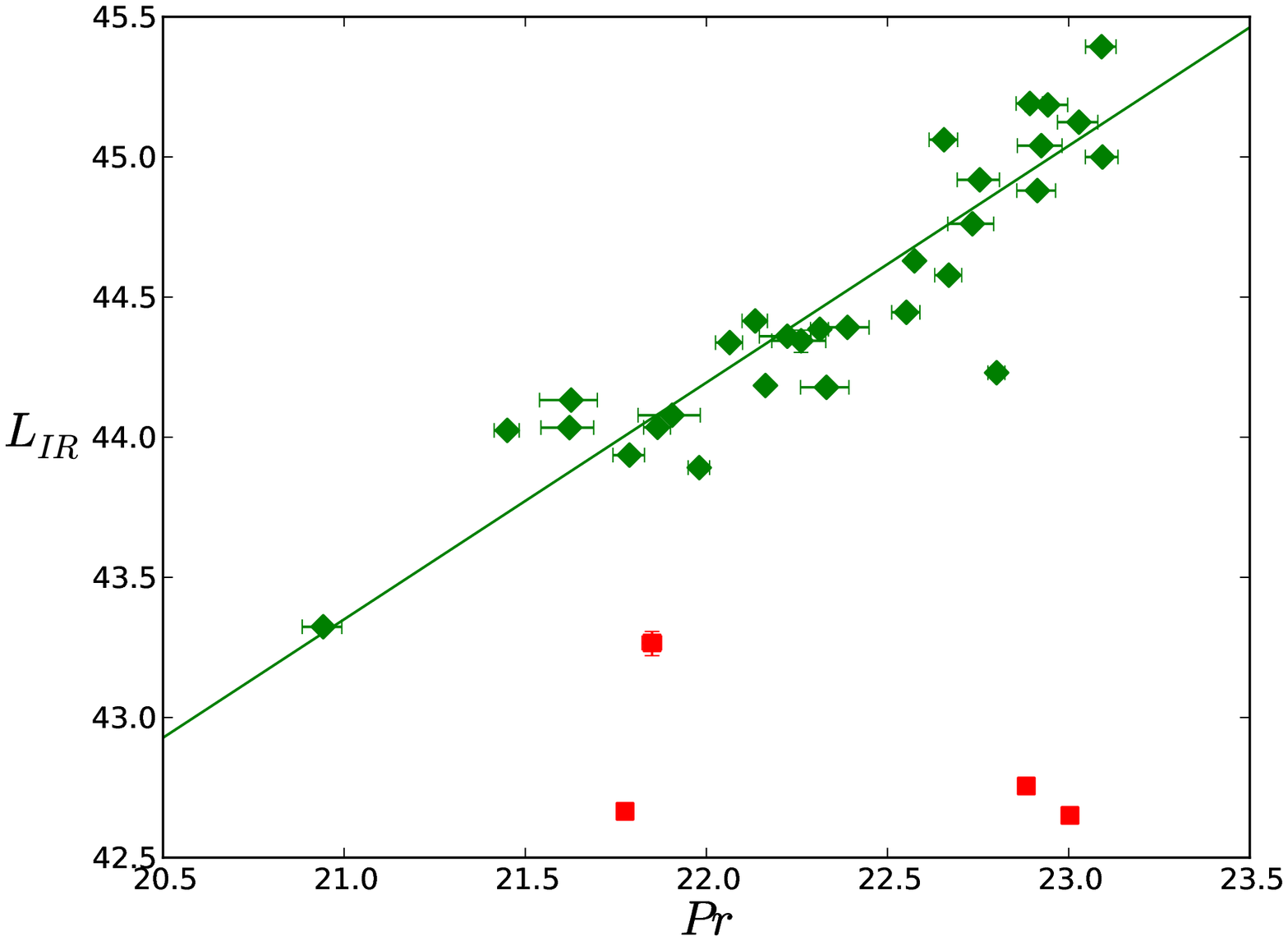} &
	\includegraphics[width=\columnwidth]{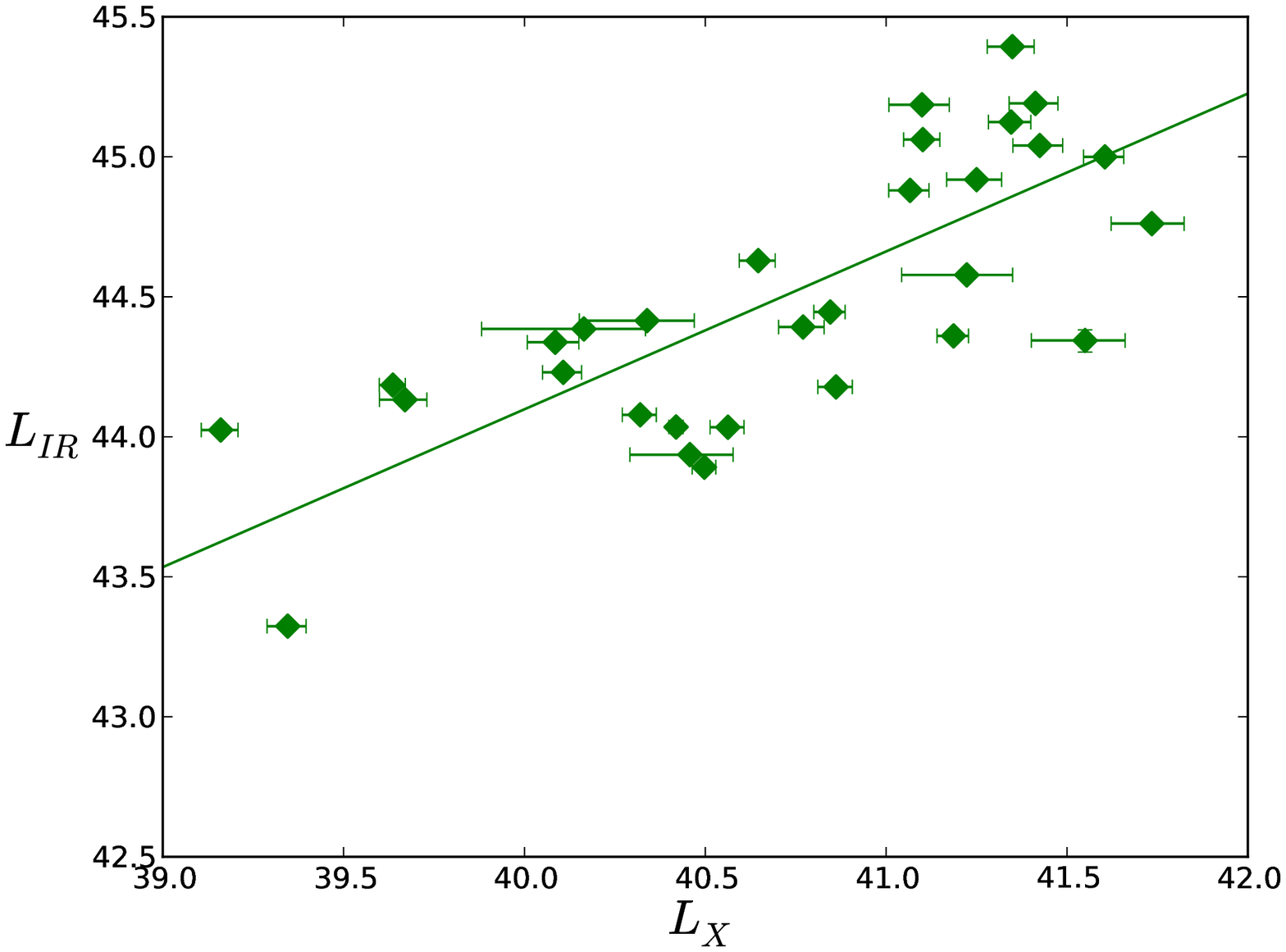} \\

	\end{tabular}
 \caption{{\sl Left}: $P_{\rm r}-L_{\rm IR}$ relation for the sub-sample of SFGs 
with X-ray counterpart selected in \citet{vattakunnel12}. The four sources classified as RL AGNs in \citet{bonzini13} are plotted as red squares.  {\sl Right}: $L_{\rm X}-L_{\rm IR}$ relation for the same subsample after removing the outliers in the $P_{\rm r}-L_{\rm IR}$ relation.}
 \label{fig_Lx_Lir_Pr}
\end{figure*}

The information on the IR properties of the E-CDFS sources presented in
this work, allow us to further explore the relation between $L_{\rm
  X}$  and SFR in this sub-sample of sources.  
As a first step, we consider only 35 sources, out of 43, which have a $>5\sigma$ radio detection, and plot the $L_{\rm FIR}$-$P_r$ relation. We
find that the bulk of the sources follow the radio-infrared
correlation expected for star forming galaxies  (see
Fig. \ref{fig_Lx_Lir_Pr}, left panel), confirming that the selection
was effective in selecting sources powered by SF processes.  However,
we find 4 clear outliers at low IR luminosities, that indeed were classified as RL AGNs according to their $q_{24obs}$ value \citep[see][for details]{bonzini13}. 
For the remaining 31 sources, we plot the $L_{\rm X}$-$L_{\rm
    FIR}$ relation (see Fig. \ref{fig_Lx_Lir_Pr}, right panel). 

We also measure the observed scatter in the
$L_{\rm X}$-$L_{\rm FIR}$  and $L_{\rm FIR}$-$P_r$ relations finding
that in the first case the dispersion is $\sim$0.29 dex, about two
times larger than the one between $L_{\rm FIR}$-$P_r$ (0.17 dex). 
%The statistical error on the X-ray luminosity account for only for $<$0.9 dex 
Even after removing the statistical error associated to the X-ray and radio luminosity, the intrinsic scatter of the $L_{\rm X}$-$L_{\rm FIR}$ relation remains two times larger than the other.
Given the robust selection, we can safely assume negligible
AGN contamination, and therefore interpret this scatter as  an
intrinsic scatter associated to the SFR  tracer. Therefore, for the
first time we show that the X-ray luminosity, as a star formation
tracer, is noisier than radio and IR luminosities.
  
The scatter can be due to an increasing X-ray emission component proportional
to the stellar mass and not to the instantaneous SFR,
analogous to what has been found in a sample of luminous
infrared galaxies by \citet{lehmer10}. Alternatively, the scatter may be due to the hard component of
a strongly absorbed AGN component.
%______________________________________________
\section{Summary}
\label{sec_summary}
In this work, we investigated the star formation properties of the
faint radio population as detected by one of the deepest 1.4 GHz
survey up-to-date conducted with the VLA in the Extended Chandra Deep
Field South. This study builds upon the results presented in
\citet{bonzini12} and \citet{bonzini13} where we have exploited the
wealth of multi-wavelength data available in this field to identify
the AGNs, further divide them into radio-loud and radio-quiet, and to
characterize the properties of the radio selected galaxies
(e.g. redshift, stellar mass). The main results of this paper are the
following:
\begin{itemize}
\item We have derived the FIR luminosity, fitting the UV-to-FIR SED of
  our VLA sources; radio selected SFGs follow the RFC up to z$\sim$3
  with a nearly constant dispersion of 0.2 dex (Sec. \ref{sec_RFC})
\item Comparing the SFR derived from the FIR luminosity and the radio
  power (Table \ref{tab_src_properties}), we show that the two SFR
  tracers are equivalently good not only in non-active SFGs but also
  for the host galaxies of RQ AGNs. This implies that the main
  contribution to the radio emission in RQ AGNs (at least at $z \sim 1.5 - 2$) 
  is associated with SF
  activity in the host rather than with radio jets (if present) powered
  by the black hole (Sec. \ref{sec_radio_in_RQ})
\item As the SFR in SFGs correlates with their stellar mass, we made
  use of the sSFR to determine the SF activity level in our sample
  (Sec. \ref{sec_sSFR}). The data are consistent with models that
  predict that the SB galaxies represent a small fraction ($\sim$8\%)
  of a mass selected SFGs population \citep[e.g.,][]{sargent12,bethermin12}.  We
  have also discussed the impact of our flux density limit on a mass
  selected mock catalogue (Sec. \ref{sec_SFinRQ}).
%Given our flux density limit of 37 $\mu$Jy, we cannot observe the bulk of the MS \citep[see][]{rodighiero11, gruppioni13}.
\item The majority of our RL AGNs are hosted in passive galaxy, but we
  detect significant SF activity in $\sim$40\% of the RL AGNs
  hosts. This suggests that, at least for low radio power, the
  presence of radio jets does not always prevent active star-formation
  (Sec. \ref{sec_SFinRL}).
\item We find hints of a higher fraction of AGNs in the most extreme
  SB galaxies (Sec. \ref{sec_AGN_content}). This is in agreement with
  scenarios where both the intense SF activity and the efficient
  accretion on the black hole are triggered by gas rich major mergers \citep[e.g.,][]{chen13,symeonidis13}.
\item The vast majority ($\sim$75\%) of our RQ AGNs % at the flux density limit of our VLA survey
 lie along the main
  sequence suggesting that the bulk of the black hole activity is
  associated with secular processes.
\end{itemize}
Finally, we have shown that deep radio continuum surveys are a powerful
tool to investigate the star formation history up to high
redshift. The current and up-coming radio facilities like the JVLA and
the Square Kilometre Array (SKA) pathfinders will be able to observe
with nanoJy ($1\sigma$) sensitivity large areas of the sky \citep{norris13} hence
detecting the bulk of the SF population.
Therefore, our finding that the main
contribution of radio emission in RQ AGNs is due to star formation in
their host galaxy opens the possibility to use radio emission to
estimate the SFR even in the host galaxy of bright quasars.

\section*{Acknowledgments}
This work is based on observations with the National Radio Astronomy
 Observatory which is a facility of the National Science Foundation
 operated under cooperative agreement by Associated Universities,
 Inc.
 
PACS has been developed by a consortium of institutes led by MPE (Germany) and including UVIE (Austria); KU Leuven, CSL, IMEC (Belgium); CEA, LAM (France); MPIA (Germany); INAF-IFSI/ OAA/OAP/OAT, LENS, SISSA (Italy); IAC (Spain). This development has been supported by the funding agen- cies BMVIT (Austria), ESA-PRODEX (Belgium), CEA/CNES (France), DLR (Germany), ASI/INAF (Italy), and CICYT/MCYT (Spain).

\bibliography{BonziniBiblio}{}
\bibliographystyle{mn2e}

%\newpage
\appendix
\section{Choice of the model parameter}
\label{sec_model_param}
We describe here in more details the empirical model used
to build the mock catalogue \citep{bernhard14} and discuss the set of model parameters
adopted. The physical motivation for the model and the underlying
equations are described in \citet{sargent12} and \citet{bethermin12}.

The motivation for this Appendix is that the original set of parameters adopted in \citet{bethermin12} needed to be modified in order to
reproduce our observations. Indeed, the corresponding mock catalogue, "observed" with the same flux density limit of our survey (see Sec. \ref{sec_model_description}), while reproducing approximatively the radio counts, does not well reproduce some host galaxy properties of the sample; the mock catalogue has a mass distribution that peaks at higher masses compared to the distribution for our VLA sources. 
%A KS-test confirms the statically significant difference between the two distribution (Prob<<1)   
As a consequence, also the SFR properties are not well reproduced in the mock catalogue having on average lower SFR and therefore a significantly lower fraction of starburst galaxies compared to the data. 

Therefore, we investigated the possible reasons for this discrepancy and looked for a set of parameters that is able to reproduce our observations in terms of mass and sSFR distribution together with the radio counts. 

\subsection{SFGs mass function}
%\subsection{Mass distribution}
The mass distribution of the mock catalogue objects is set by the SFG
mass function (MF). The MF adopted in the model is based on the fits
by \citet{peng10} of the SFGs MF presented in \citep{ilbert10}.  It is
described as a single Schechter function with characteristic mass
($M_b$) and faint-end slope that are redshift invariant and by a
constant characteristic density up to z$\sim$1 followed by a decline
as $(1-z)^{0.46}$ (see \citet{bethermin12} for details).  Our radio
observations are relatively shallow compared to the K-band data used
to compute the MF in \citet{ilbert10}, consequently we are more
sensitive to the high mass end of the distribution. Therefore, we are
particularly sensitive to the value of $M_b$, that set the position of
the MF break followed by the exponential cut off. Since our sample is
small and highly incomplete at low masses, we cannot perform a real
fit of the MF. We chose to keep all the other parameter describing the
MF fixed.  As already mentioned our observations have a mass
distribution that peaks at lower masses compared to the corresponding
mock catalogue.  To reconcile the model with our observations we assume
a $M_b$ of $10^{11} M_{\odot}$ rather than $10^{11.2} M_{\odot}$ as in
\citet{bethermin12}.  Adopting this value the mass distributions of
the mock catalogue and of our data are consistent as confirmed by a
Kolmorogov-Smirnov (KS)-test (Prob$>0.99$).  The possible reasons for
this shift are twofold: on one hand, stellar mass measurements have
large uncertainties and can be up to 0.3 dex systematically different
depending on the stellar population synthesis model adopted \citep[see
  e.g.][]{ilbert10}.
To check this hypothesis we computed the stellar mass using the same
method adopted for the VLA sources \citep{bonzini13} for a sub-sample
of the \citet{ilbert10} sources\footnote{A sample of X-ray selected
  AGNs for which we have optical-to-24$\mu$m photometry and redshift.}
and compared the mass measurements. We indeed noted that the
\citet{ilbert10} stellar masses are on average larger than the one
obtained with our method.  On the other hand, there is no one-to-one
correspondence between the color-color based method used to identify
the SFG population in \citet{ilbert10} and our scheme; it is therefore possible that some sources excluded from the \citet{ilbert10} SFG sample because of their redder colours have been considered SFGs according to our classification.
Finally, we note that the simple prescription adopted in the empirical
model and the set of parameters chosen is only an approximation for
the SFGs MF while a more complex description as e.g. in
\citet{ilbert13} would be needed.

\subsection{Main sequence and its redshift evolution}
The two modes of SF are described by two Gaussians \citep{sargent12}. We assume a FWHM of
0.2 dex for the MS galaxies as measured in \citet{rodighiero11}. For
the second Gaussian we use the same parameter adopted in
\citep{bethermin12}, i.e. a FHWM of 0.2 dex, a displacement from the
MS peak of 0.6 dex. The relative fraction of SB evolves with redshift
as:
\begin{equation}
f_{SB}= 0.012\times (1+z)^2\ for\ z<1
\end{equation} 
The local fraction is taken from \citet{bethermin12} but we set a
steeper growth of the starburst fraction up to redshift 1 as suggested
by our observations (see Sec. \ref{sec_SFinRQ}).
%and as found in \citet{gruppioni13}

The normalization of the MS is another critical parameter in
determining both the number counts and the observed SB fraction.
Keeping fixed the slope of the MS to the value measured in
\citet{rodighiero11} and its evolution with redshift to $(1+z)^{2.8}$
\citep{sargent12}, we looked for the best value for the MS
normalization.  We adopt a normalization of the MS ($logsSFR(z=0,
M=10^{11} M_{\odot})= 10.08$) that is lower than in \citet{bethermin10}
but it is in better agreement with what was found in \citet{rodighiero11}
and our own measurement. It is also consistent with previous results in the literature up to $z \sim$2 \citep{noeske07,elbaz07,daddi07a}.
Recent work suggests that above $z \sim$2 the evolution of the MS is shallower than $(1+z)^{2.8}$ as assumed by the \citet{bernhard14} model \citep[e.g.][]{weinmann11,stark13,gonzalez14}. As a consequence, the MS shown in the last panel of Fig. \ref{fig_SFR-Mstar} is probably too high. However, only about 5\% of our sources fall in this high redshift bin (2.5$< z < $4) and therefore the simplification adopted in the model does not affect significantly our conclusions. We stress that the main goal of the comparison of our data with the mock catalogue was to better quantify the selection effects in flux density limited samples, to check the qualitative agreement with the two Gaussian SF mode model \citep{sargent12} and not to obtain a fit of the MS as a function of redshift.

\subsection{Comparison with VLA observations}
With the few changes in the model parameters described above, we are
able to reproduce both the radio counts and the physical properties of
our radio sources with the mock catalogue.  Indeed the mass selected
mock catalogue, "observed" with the same flux density limit of our VLA
survey (see Sec.\ref{sec_model_description}), on a mock field with the
same area of our observation, contains about 790 sources.  The exact
number of sources varies of some units for different runs of the model
since we add random uncertainties on the mock galaxies properties and
a random dispersion for the RFC. The VLA sample considered in this
work contains 779 sources. Considering only the RQ AGNs and the SFGs,
where the main contribution to the radio flux is due to SF, but
correcting for the not perfect uniformity of the sky coverage
\citep{padovani15}, the number of RQ AGNs and SFGs detectable
at the 37$\mu$Jy flux density limit in our field is 784.  Considering that
some of the RL AGNs, according to our analysis, contribute to the SFG
population, we conclude that the number counts predicted by the
empirical model are in good agreement with the observed one. A
complete analysis of the impact of the various model parameters on the
mock catalogue physical characteristics is beyond the scope of this
paper.
The goal of this investigation was to check the consistency between
our observation and the model for the SF population proposed in
\citet{sargent12} and to better control the effects of the flux
density limit on our results.

\section{Stellar masses and SFRs catalogue of VLA sources}
\label{sec_cat}
We make publicly available the physical properties derived for our
radio sample and used in this work.  They are summarized in table
\ref{tab_src_properties}.  The catalogue columns are organized as
follows:
\begin{itemize}
\item (1) Identification number of the radio source (RID).
\item (2) Source classification.
\item (3) Source activity: "SB" for starburst galaxies, "MS" for main sequence galaxies, and "P" for passive galaxies according to the definition given in section \ref{sec_sSFR}.
\item (4) Source redshift.
\item (5) Stellar mass.
\item (6) SFR derived from the radio power.
\item (7) SFR derived from the FIR luminosity.
\item (8) distance with respect to the main sequence in the $SFR-M_{star}$ plane  (Sec. \ref{sec_AGN_content})
\item (9) PACS flag; 0= only upper limits in the PACS bands, 1= at least one PACS detection
\end{itemize}

\begin{table*}

\caption{Star formation properties of the VLA sources.(The full
Table is available in the on-line journal. A portion in shown here for guidance regarding its form and content.)}
\begin{tabular}{r c c c c c c c c}
\hline
  (1) & (2) & (3) & (4) & (5) & (6) & (7) & (8)  & (9) \\
id & class & type & z & $M_{star}$ & $SFR_{\rm r}$ & $SFR_{\rm FIR}$ & $sSFRex$ & PACS? \\
       &            &        & & [$log (M_{\odot})$] & [$log (M_{\odot} yr^{-1})$] & [$log (M_{\odot} yr^{-1})$] & & \\
\hline
711 & RL AGN & P & 1.89 & 10.77$\pm$0.18 & 277.9$\pm$77.6 & 4$\pm$0.1 & -1.30 & 0 \\
712 & SFG & MS & 0.56 & 10.79$\pm$0.05 & 15.4$\pm$4.7 & 21$\pm$1.1 & 0.07 & 1 \\
713 & RQ AGN & MS & 0.49 & 10.60$\pm$0.06 & 17.2$\pm$3.3 & 10$\pm$0.8 & -0.01 & 1 \\
714 & SFG & MS & 0.77 & 10.30$\pm$0.02 & 41.0$\pm$9.3 & 30$\pm$4.0 & 0.47 & 1 \\
715 & SFG & SB & 0.18 & 9.95$\pm$0.07 & 13.3$\pm$0.8 & 11$\pm$0.2 & 0.81 & 1 \\
716 & RQ AGN & MS & 1.16 & 10.78$\pm$0.10 & 80.2$\pm$25.1 & 79$\pm$20.9 & 0.27 & 1 \\
717 & SFG & MS & 0.52 & 10.13$\pm$0.02 & 17.7$\pm$3.8 & 12$\pm$0.9 & 0.41 & 1 \\
718 & SFG & SB & 2.46 & 10.52$\pm$0.09 & 478.7$\pm$141.2 & 703$\pm$86.2 & 0.84 & 1 \\
719 & RL AGN & MS & 1.03 & 10.84$\pm$0.19 & 38394.1$\pm$20.2 & 19$\pm$0.2 & -0.31 & 0 \\
720 & SFG & MS & 0.25 & 10.59$\pm$0.02 & 16.5$\pm$2.4 & 10$\pm$0.5 & 0.19 & 1 \\
\hline
\end{tabular}
\label{tab_src_properties}
\end{table*}

%---------------------------------------------------------
\clearpage

\label{lastpage}

%\bsp

\end{document}